\documentclass[11pt]{article}
\pdfoutput = 1
\usepackage{float}
\usepackage{graphicx,color}
\usepackage{epsfig}
\usepackage{amsmath,amssymb,amsthm}
\usepackage{subfigure}
\usepackage{dsfont}
\usepackage{psfrag}
\usepackage{algpseudocode}

\def\1{\mathbf{1}}

\graphicspath{{figures/}}

\newtheorem{theorem}{Theorem}[section]

\newtheorem{proposition}[theorem]{Proposition}

\newtheorem{remark}[theorem]{Remark}

\floatstyle{ruled}
\newfloat{Algorithm}{tbp}{lop}[section]

\providecommand{\norm}[1]{\lVert#1\rVert}

\begin{document}

\title{Iterative Methods for Scalable Uncertainty Quantification in Complex Networks}

\author{Amit Surana, Tuhin Sahai and Andrzej Banaszuk}
\date{United Technologies Research Center, 411 Silver Lane, East Hartford, CT 06108, USA}
\maketitle

\begin{abstract}
In this paper we address the problem of uncertainty management for robust design, and verification of large dynamic networks whose performance is affected by an equally large number of uncertain parameters. Many such networks (e.g. power, thermal and communication networks) are often composed of weakly interacting subnetworks. We propose intrusive and non-intrusive iterative schemes that exploit such weak interconnections to overcome dimensionality curse associated with traditional uncertainty quantification methods (e.g. generalized Polynomial Chaos, Probabilistic Collocation) and accelerate uncertainty propagation in systems with large number of uncertain parameters. This approach relies on integrating graph theoretic methods and waveform relaxation with generalized Polynomial Chaos, and Probabilistic Collocation, rendering these techniques scalable. We analyze convergence properties of this scheme and illustrate it on several examples.
\end{abstract}

\section{Introduction}
The issue of management of uncertainty for robust system operation is of interest in a large family of complex networked systems. Examples include power, thermal and communication networks which arise in several instances such as more electric aircrafts, integrated building systems and sensor networks. Such systems typically involve a large number of heterogeneous, connected components, whose dynamics is affected by possibly an equally large number of uncertain parameters and disturbances.

Uncertainty Quantification (UQ) methods provide means of calculating probability distribution of system outputs, given probability distribution of input parameters. Outputs of interest could include for example, latency in communication network, power quality and stability of power networks, and energy usage in thermal networks. The standard UQ methods such as Monte Carlo(MC) \cite{MCS} either exhibit poor convergence rates or others such as Quasi Monte Carlo (QMC) \cite{nets}\cite{lattice}, generalized Polynomial Chaos(PC) \cite{gPC} and the associated Probabilistic Collocation method (PCM) \cite{xiu}, suffer from the curse of dimensionality (in parameter space), and become practically infeasible when applied to network as a whole. Improving these techniques to alleviate the curse of dimensionality is an active area of current research (see \cite{cursedim} and references therein): notable methods include sparse-grid collocation method \cite{sparse},\cite{MEPCM} and  ANOVA decomposition \cite{Anova} for sensitivity analysis and dimensional reduction of the uncertain parametric space. However, none of such extension exploits the underlying structure and dynamics of the networked systems. In fact, many networks of interest (e.g. power, thermal and communication networks), are often composed of weakly interacting subsystems. As a result, it is plausible to simplify and accelerate the simulation, analysis and uncertainty propagation in such systems by suitably decomposing them. For example, authors in \cite{gp1,gp2} used graph decomposition to facilitate stability and robustness analysis of large-scale interconnected dynamical systems. Mezic et al. \cite{igorcdc} used graph decomposition in conjunction with Perron Frobenius operator theory to simplify the invariant measure computation and uncertainty quantification, for a particular class of networks. While these approaches exploit the underlying structure of the system, they do not take advantage of the weakly coupled dynamics of the subsystems.

In this paper, we propose an iterative UQ approach that exploits the weak interactions among subsystems in a networked system to overcome the dimensionality curse associated with traditional UQ methods. We refer to this approach as Probabilistic Waveform Relaxation (PWR), and propose both intrusive and non-intrusive forms of PWR. PWR relies on integrating graph decomposition techniques and waveform relaxation scheme, with gPC and PCM. Graph decomposition to identify weakly interacting subsystems, can be realized by spectral graph theoretic techniques \cite{Tutorial},\cite{Chung}. Waveform relaxation \cite{wave} (WR), a parallelizable iterative method, on the other hand, exploits this decomposition and evolves each subsystem forward in time independently but coupled with the other subsystems through their solutions from the previous iteration. In the intrusive PWR, the subsystems obtained from decomposing the original system are used to impose a decomposition on system obtained by Galerkin projection based on the gPC expansion. Further the weak interactions are used to discard terms which are expected to be insignificant in the gPC expansion, leading to what we call an Approximate Galerkin Projected (AGP) system. We then propose to apply WR relaxation on the decomposed AGP system to accelerate the UQ computation. In the non-intrusive form of PWR, rather than deriving the AGP system, one works directly with subsystems obtained from decomposing the original system. At each waveform relaxation iteration we propose to apply PCM at subsystem level, and use gPC to propagate the uncertainty among the subsystems. Since UQ methods are applied to relatively simpler subsystems which typically involve a few parameters, this renders a scalable  non-intrusive iterative approach to UQ. We prove convergence of the PWR approach under very general conditions. Note that spectral graph decomposition can be done completely in a distributed fashion using a recently developed wave equation based clustering method \cite{ref:wave}. Moreover, one can further exploit timescale separation in the system to accelerate WR using an adaptive form of WR \cite{AWR}.  PWR when combined with wave equation based distributed clustering and adaptive WR can lead to highly scalable and computationally efficient approach to UQ in complex networks.

This paper is organized in six sections.  In section \ref{UQ} we give the mathematical preliminaries for setting up the UQ problem for networked dynamical systems,  and present an overview of gPC and PCM techniques. In section \ref{nd} we discuss graph decomposition and waveform relaxation methods, which form basic ingredients of PWR. Here we also describe  adaptive WR and wave equation based distributed graph decomposition techniques. We introduce the intrusive and non-intrusive PWR in section \ref{scaleUQ} through a simple example, and then describe these methods in a more general setting. We also prove convergence of PWR, and analyze the scalability of the method. In section \ref{examples} we illustrate the intrusive and non-intrusive PWR on several examples. Finally, in section \ref{conc} we summarize the main results of this paper, and present some future research directions.

\section{Uncertainty Quantification in Networked Systems}\label{UQ}
Consider a nonlinear system described by a system of random differential equation
\begin{eqnarray}
  \dot{x}_1&=&f_1(\mathbf{x},\mathbf{\xi}_1,t),\notag\\
    \vdots\notag\\
  \dot{x}_n&=&f_n(\mathbf{x},\mathbf{\xi}_n,t),\label{complexsys}
\end{eqnarray}
where, $\mathbf{f}=(f_1,f_2,\cdots,f_n)\in\mathbb{R}^n$ is a smooth vector field, $\mathbf{x}=(x_1,x_2,\cdots,x_n)\in \mathbb{R}^n$ are state variables, $\mathbf{\xi}_i\in R^{p_i}$ is vector of random variables affecting the $i-$th system. Let $\mathbf{\xi}=(\mathbf{\xi}_1^T,\cdots,\mathbf{\xi}_n^T)^T\in \mathbb{R}^p$ be the $p=\sum_{i=1}^n p_i$ dimensional random vector of uncertain parameters affecting the complete system. The solution to initial value problem $\mathbf{x}(t_0)=\mathbf{x}_0$ will be denoted by $\mathbf{x}(t;\mathbf{\xi})$, where for brevity we have suppressed the dependence of solution on initial time $t_0$ and initial condition $\mathbf{x}_0$. We shall assume that the system (\ref{complexsys}), is Lipschitz
\begin{equation}\label{LipF}
||\mathbf{f}(\mathbf{x}_1,\xi,t)-\mathbf{f}(\mathbf{x}_2,\xi,t)||\leq L(\xi)||\mathbf{x}_1-\mathbf{x}_2||,
\end{equation}
where, the Lipschitz constant $L(\xi)$ depends on the random parameter vector and $||\cdot||$ is a Euclidean norm. We will assume that $\sup_{\mathbf{\xi}\in  \mathbb{R}^p} L(\xi)=L<\infty$.

Let us also define a set of quantities
\begin{equation}\label{observe1}
\mathbf{z}=(z_1,z_2,\cdots,z_d)=G(\mathbf{x})=(g_1(\mathbf{x}),\cdots,g_d(\mathbf{x})),
\end{equation}
as observables or quantities of interests. The goal is to numerically establish the effect of input uncertainty of $\mathbf{\xi}$ on output observables $\mathbf{z}$. Naturally, the solution for system (\ref{complexsys}) and the observables (\ref{observe1}) are functions of same set of random variables $\mathbf{\xi}$, i.e
\begin{equation}\label{observe2}
\mathbf{x}=\mathbf{x}(t;\mathbf{\xi}),\qquad \mathbf{z}=\mathbf{z}(t,\xi)=G(\mathbf{x}).
\end{equation}

In what follows we will adopt a probabilistic framework and model  $\mathbf{\xi}=(\xi_1,\xi_2,\cdots,\xi_p)$ as a $p-$ variate random vector with independent components in the probability space $(\Omega,\mathcal{A},\mathcal{P})$, whose event space is $\Omega$ and is equipped with $\sigma-$algebra $\mathcal{A}$ and probability measure $\mathcal{P}$. Throughout this paper, we will assume that the parameters $\Sigma=\{\xi_1,\xi_2,\cdots,\xi_p\}$ are mutually independent of each other. Let $\rho_i:\Gamma_i\rightarrow \mathbb{R}^+$ be the probability density of the random variable $\xi_i(\omega)$, with $\Gamma_i=\xi_i(\Omega)\subset \mathbb{R}$ being its image. Then, the joint probability density of
any random parameter subset $\Lambda=\{\xi_{i_1},\xi_{i_2},\cdots,\xi_{i_m}\}\subset \Sigma$
is given by
\begin{equation}\label{jointpdfsub}
\mathbf{\rho}_{\Lambda}(\xi_{i_1},\cdots,\xi_{i_m})=\prod_{j=1}^{|\Lambda|}\rho_{i_j}(\xi_{i_j}),\quad \forall (\xi_{i_1},\cdots,\xi_{i_m})\in \Gamma_{\Lambda},
\end{equation}
with a support
\begin{equation}\label{jointspace}
\Gamma_{\Lambda}=\prod_{j=1}^{|\Lambda|}\Gamma_{i_j}\subset \mathbb{R}^{|\Lambda|},
\end{equation}
where, $|\cdot|$ denotes the cardinality of the set. Without loss of generality we will assume that $\Gamma_i=[-1\quad 1],i=1,\cdots,p$.

\begin{remark}
While throughout the paper we will work with ODEs (\ref{complexsys}) with parametric uncertainty, the PWR framework developed in this paper can be naturally extended to deal with 1) System of differential algebraic equations (DAEs), and 2) Time varying uncertainty. Both these extensions would be illustrated through examples in section \ref{examples}.
\end{remark}

\subsection{Uncertainty Quantification Methods}
In this section, we describe two interrelated UQ approaches: generalized polynomial chaos (gPC) and probabilistic collocation method (PCM). The gPC is an intrusive approach which requires explicit access to system  equations (\ref{complexsys}), while PCM is a related sampling based non-intrusive (and hence treats the system (\ref{complexsys}) as a black box) way of implementing gPC.

\subsubsection{Generalized Polynomial Chaos}\label{gPC}
In the finite dimensional random space $\Gamma_{\Sigma}$ defined in (\ref{jointspace}), the gPC expansion seeks to approximate a random process via orthogonal polynomials of random variables. Let us define one-dimensional orthogonal polynomial space
associated with each random variable $\xi_k, k=1,\cdots,p$ as
\begin{equation}\label{polyspacegenloc}
W^{k,d_k}\equiv\{v:\Gamma_k\rightarrow \mathbb{R}:v\in\mbox{span}\{\psi_i(\xi_k)\}_{i=0}^{d_k}\},
\end{equation}
where, $\{\psi_i(\xi_k)\}_{i=0}^{d_k}$ denotes the orthonormal polynomial basis from the so called Wiener-Askey polynomial chaos \cite{gPC}. The Askey scheme of polynomials contains various classes of orthogonal polynomials such that their associated weighting functions coincide with probability density function of the underlying random variables. The corresponding $P$-variate orthogonal polynomial space in $\Gamma_{\Sigma}$ is defined as
\begin{equation}\label{polyspacegen}
W^{\Sigma,P}\equiv \bigotimes_{|\mathbf{d}|\in \mathbb{P}}W^{i,d_i},
\end{equation}
where the tensor product is over all possible combinations of the multi-index $\mathbf{d}=(d_1,d_2,\cdots,d_{|\Sigma|}) \in \mathbb{N}^{|\Sigma|}$ in set $\mathbb{P}$,
\begin{equation}\label{Pdef}
\mathbb{P}=\{\mathbf{d}\in \mathbb{N}^{|\Sigma|} : |\mathbf{d}|=\sum_{i=1}^{|\Sigma|} d_i\leq \overline{P} \quad \mbox{and} \quad d_i\leq P_i\}
\end{equation}
and, $P=(P_1,\cdots,P_{|\Sigma|})^T\in \mathbb{N}^{|\Sigma|}$ is vector of integers which restricts the maximum order of expansion of $i$-th variable $\xi_i$ to be $P_i$, and $\overline{P}=\max_{i}P_i$.
Thus, $W^{\Sigma,P}$ is the space of N-variate orthonormal polynomials of total degree at most $\overline{P}$ with additional constraints on individual degrees of polynomials, and its basis functions $\Psi_i^{\Sigma,P}(\mathbf{\xi})$ satisfy
\begin{equation}\label{orthogen}
\int_{\Gamma_{\Sigma}}\Psi_i^{\Sigma,P}(\mathbf{\xi})\Psi_j^{\Sigma,P}(\mathbf{\xi})\mathbf{\rho}_{\Sigma}(\xi)d\mathbf{\xi}=\delta_{ij},
\end{equation}
for all $1\leq i,j\leq N_{\Sigma}=\mbox{dim}(W^{\Sigma,P})$. Note that in standard gPC all expansion orders are taken to be identical i.e. $P_1,=P_2\cdots=P_{|\Sigma|}=\overline{P}$, so that $\mbox{dim}(W^{\Sigma,p})=\frac{(\overline{P}+|\Sigma|)!}{\overline{P}!|\Sigma|!}$. We will however take advantage of an adaptive expansion, a notion which will be fully developed in section \ref{approxPWR}.

The major advantage of applying the gPC is that a random differential equation can be transformed into a system of deterministic equations. A typical approach is to employ a stochastic Galerkin projection, in which all the state variables are expanded in polynomial chaos basis with corresponding modal coefficients ($a_{ik}(t)$), as
\begin{equation}\label{expx}
x_k(t,\mathbf{\xi})\approx x_k^{\Sigma,P}(t,\mathbf{\xi})=\sum_{i=1}^{N_{\Sigma}} a_{ik}(t)\Psi_i^{\Sigma,P}(\mathbf{\xi}),\quad k=1,\cdots,n,
\end{equation}
where, the sum has been truncated to a finite order. Substituting, these expansions in Eq. (\ref{complexsys}),  and using the orthogonality property of polynomial chaos (\ref{orthogen}), we obtain for $k=1,\cdots,n,\quad j=1,\cdots,N_{\Sigma}$,
\begin{equation}\label{galproj}
\dot{a}_{jk}=F_{jk}(\mathbf{a},t),
\end{equation}
a system of deterministic ODE�s describing the evolution of the modal coefficients, with initial conditions
\begin{equation}\label{galprojinit}
a_{jk}(0)=\int_{\Gamma_{\Sigma}}x_k(0,\mathbf{\xi})\Psi_j^{\Sigma,P}(\mathbf{\xi})\mathbf{\rho}_{\Sigma}(\mathbf{\xi})d\mathbf{\xi},
\end{equation}
and $\mathbf{a}=(a_{11},\cdots,a_{N_{\Sigma}1},\cdots,a_{1n},\cdots,a_{N_{\Sigma}n})^T$,
\begin{equation}\label{Fdef}
F_{jk}(\mathbf{a},t)=\int_{\Gamma_{\Sigma}}f_k(\mathbf{x}^{\Sigma,P}(\mathbf{\xi},t),\mathbf{\xi}_k,t)\Psi_j^{\Sigma,P}(\mathbf{\xi})\mathbf{\rho}_{\Sigma}(\mathbf{\xi})d\mathbf{\xi},
\end{equation}
with $\mathbf{x}^{\Sigma,P}(t,\mathbf{\xi})=(x_1^{\Sigma,P}(t,\mathbf{\xi}),\cdots,x_n^{\Sigma,P}(t,\mathbf{\xi}))$.
This system can be solved with any numerical method dealing with initial-value problems, e.g., the Runge-Kutta method. Similarly, the observable can be expanded in gPC basis, as
\begin{equation}\label{expz}
z_k(t,\mathbf{\xi})\approx z_k^{\Sigma,P}(t,\mathbf{\xi})=\sum_{i=1}^{N_{\Sigma}} b_{ik}(t)\Psi_i^{\Sigma,P}(\mathbf{\xi}),
\end{equation}
where,
\begin{eqnarray}
b_{jk}(t)&=&\int_{\Gamma_{\Sigma}}z_k^{\Sigma,P}(\mathbf{\xi})\Psi_j^{\Sigma,P}(\mathbf{\xi})\mathbf{\rho}_{\Sigma}(\mathbf{\xi})d\mathbf{\xi}\label{z1approx}
\end{eqnarray}
with $\quad k=1,\cdots,d$. Hence, once the solution to the system (\ref{galproj}) has been obtained, the coefficients $b_{jk}$ can be approximated as
\begin{equation}
b_{jk}\approx \int_{\Gamma_{\Sigma}}g_k(\mathbf{x}^{\Sigma,P}(t,\mathbf{\xi}))\Psi_j^{\Sigma,P}(\mathbf{\xi})\mathbf{\rho}_{\Sigma}(\mathbf{\xi})d\mathbf{\xi}.
\end{equation}
Such Galerkin procedures have been used extensively in the literature. In many instances Galerkin projection may not be possible due to unavailability of
direct access to the system equations (\ref{complexsys}). In many other instances such
intrusive methods are not feasible even in cases when the system equations are available, because of the cost of deriving
and implementing a Galerkin system within available computational tools. To circumvent this difficulty, probabilistic collocation method has been developed.

\subsubsection{Probabilistic Collocation Method}\label{PCM}
 PCM is a non-intrusive approach to solving stochastic random processes with the gPC. Instead of projecting each state variable onto the polynomial chaos basis, the collocation approach evaluates the integrals of form (\ref{z1approx}) by evaluating integrand at the roots of the appropriate basis polynomials \cite{xiu}.  Given a 1D probability density function $\rho_j(\xi_j)$, the PCM based on Gauss quadrature rule, approximates an integral of a function $g$ with respect to density $\rho_j(\xi_j)$, as follows
\begin{equation}\label{1D}
\int_{-1}^1g(\xi_j)\rho(\xi_j)d\xi_j\approx\mathcal{U}_{l_j}[g]=\sum_{k=1}^{m_{l_j}}w_{l_jk}g(r_{l_j k}), \quad j=1,\cdots,p,
\end{equation}
where, $r_{l_j k}\in C_{l_j}$ is the set of Gauss collocation points with associated weights $w_{l_jk}$, $l_j$ is the accuracy level of quadrature formula, and $m_{l_j}$ is the number of quadrature points corresponding to this accuracy level. Building on the 1D quadrature formula,  the full grid PCM leads to following cubature rule,
\begin{eqnarray}
&&\int_{-1}^1\int_{-1}^1\cdots \int_{-1}^1 g(\xi_1,\cdots,\xi_p)\mathbf{\rho}_{\Sigma}(\mathbf{\xi})d\mathbf{\xi}\notag\\
&\approx& \mathcal{I}(l_1,\cdots,l_p,p)[g]=(\mathcal{U}_{l_1}\otimes\mathcal{U}_{l_2}\cdots\mathcal{U}_{l_p})[g]\notag\\
&=&\sum_{j_1=1}^{m_{l_1}}\cdots \sum_{j_p=1}^{m_{l_p}}w_{\mathbf{l}\mathbf{j}}g(\mathbf{r}_{\mathbf{l}\mathbf{j}}),\label{ndint}
\end{eqnarray}
where, $\mathbf{r}_{\mathbf{l}\mathbf{j}}=(r_{l_1j_1},\cdots,r_{l_pj_p})$, with $\mathbf{l}=(l_1,\cdots,l_p)$, and $\mathbf{j}=(j_1,\cdots,j_p)$ and $w_{\mathbf{l}\mathbf{j}}=w_{l_1j_1}\cdots w_{l_nj_n}$.
To compute $\mathcal{I}(\mathbf{l},p)$ we need to evaluate the function on the full collocation  grid $C(\mathbf{l},p)$ which is given by tensor product of 1D grids
\begin{equation}\label{fullgird}
\mathcal{C}(\mathbf{l},p)=C_{l_1}\times\cdots \times C_{l_p},
\end{equation}
with a total number of collocation points being $Q=\prod_{j=1}^p m_{l_j}$.  In this framework, therefore, for any $t$, the approximations to the model coefficients $a_{jk}(t)$ (see Eq. (\ref{expx})) and $b_{jk}(t)$ (see Eq. (\ref{expz})) can be obtained as
\begin{eqnarray}
a_{jk}(t)&=&\int_{\Gamma_{\Sigma}}x_k^{\Sigma,P}(t,\mathbf{\xi})\Psi_j^{\Sigma,P}(\mathbf{\xi})\mathbf{\rho}_{\Sigma}(\mathbf{\xi})d\mathbf{\xi}\notag\\
&\approx&\sum_{j_1=1}^{m_{l_1}}\cdots \sum_{j_p=1}^{m_{l_p}}w_{\mathbf{l}\mathbf{j}}\Psi_j^{\Sigma,P}(\mathbf{r}_{\mathbf{l}\mathbf{j}})x_k(t,\mathbf{r}_{\mathbf{l}\mathbf{j}}), \label{z4approx}
\end{eqnarray}
with similar expression for $b_{jk}(t)$.
Note to compute summations arising in (\ref{z4approx}), the solution  $\mathbf{x}(t,\mathbf{r}_{\mathbf{l}\mathbf{j}})$ of the system (\ref{complexsys}) is required for each collocation point $\mathbf{r}_{\mathbf{l}\mathbf{j}}$ in the full collocation  grid  $C(\mathbf{l},p)$. Thus, simplicity of collocation framework only requires repeated runs of deterministic solvers, without explicitly requiring the projection step in gPC.

If we choose the same order of collocation points in each dimension, i.e. $m_{l_1}=m_{l_2},\cdots=m_{l_p}\equiv l$, the total number of points is $Q=l^p$, and the computational cost increases rather steeply with the number of uncertain parameters $p$. Hence, for large systems ($n \gg 1$) with large number of uncertain parameters ($p\gg1$), PCM becomes computationally intensive. As discussed in the introduction, alleviating this curse of dimensionality is an active area of current research \cite{cursedim}. In this paper we propose a new uncertainty quantification approach which exploits the underlying network structure and dynamics to overcome the dimensionality curse associated with PCM. The key methodologies for accomplishing this are the graph decomposition and waveform relaxation, which are discussed in subsequent sections.

\section{Graph Decomposition and Waveform Relaxation}\label{nd}

\subsection{Waveform Relaxation}\label{wave}
In this section we describe the basic mathematical concept of the Waveform Relaxation (WR) method for iteratively solving the system of differential equations of the form (\ref{complexsys}). For purposes of discussion here, we fix the parameter values $\mathbf{\xi}$ in the system (\ref{complexsys}) to fixed mean values.
The general structure of a WR algorithm for analyzing system (\ref{complexsys}) over a given time interval $[0, T]$ consists of two major steps: \emph{assignment partitioning process} and the \emph{relaxation process} \cite{waveform1,wave}.

\emph{Assignment-partitioning process}: Let $\mathcal{N}=\{1,\cdots,n\}$ be the set of state indices, and $\mathcal{C}_i,i=1,\cdots,m$ be a partition of $\mathcal{N}$ such that
\begin{equation}\label{part}
\mathcal{N}=\bigcup_{i=1}^m\mathcal{C}_i,\qquad \mathcal{C}_i\bigcap \mathcal{C}_j=\emptyset,\forall i\neq j.
\end{equation}
We shall represent by $\mathcal{D}:\mathcal{N}\rightarrow \mathcal{M}\equiv\{1,2,\cdots,m\}$ a map which assigns the state index to its partition index, i.e. $\mathcal{D}(i)=j$ where, $j$ is such that $i\in \mathcal{C}_j$. Without loss of generality, we can rewrite Eq. \ref{complexsys} after the assignment-partitioning process as:
\begin{eqnarray}
  \dot{\mathbf{y}}_1 &=&\mathbf{F}_1(\mathbf{y}_1,\mathbf{d}_1(t),\Lambda_1,t)\notag \\
  \vdots\notag\\
  \dot{\mathbf{y}}_m &=&\mathbf{F}_m(\mathbf{y}_m,\mathbf{d}_m(t),\Lambda_m,t),\label{decomp}
\end{eqnarray}
where, for each $i=1,\cdots,m$,
\begin{equation}\label{Fdef}
\mathbf{F}_i\equiv (f_{j^i_1},\cdots,f_{j^i_{M_i}})^T,
\end{equation}
\begin{equation}\label{ydef}
\mathbf{y}_i\equiv(x_{j^i_1},\cdots,x_{j^i_{M_i}})^T,
\end{equation}
with initial condition
\begin{equation}\label{icfory}
\mathbf{y}_{0i}\equiv(x_{0j^i_1},\cdots,x_{0j^i_{M_i}})^T,
\end{equation}
and
\begin{equation}\label{lamdef}
\Lambda_i\equiv (\mathbf{\xi}_{j^i_1},\cdots,\mathbf{\xi}_{j^i_{M_i}})^T,
\end{equation}
are the subvectors  assigned to the $i-$th partitioned subsystem, such that $ j^i_k\in \mathcal{C}_i,\quad k=1,\cdots,M_i=|\mathcal{C}_i|$ and
\begin{equation}\label{ddef}
\mathbf{d}_i(t)\equiv(\mathbf{y}_{j_i}^T,\cdots,\mathbf{y}_{j_{N_i}}^T)^T,
\end{equation}
is a decoupling vector, with $j_k\in \mathcal{M}_i$ and $k=1,\cdots,N_i=|\mathcal{M}_i|$. Here, $\mathcal{M}_i$ is the set of indices of the partitions (or subsystems) with which the $i-$th partition (or subsystem) interacts, and is given by
\begin{equation}\label{neigh}
\mathcal{N}_i=\mathcal{M}\setminus \Im(\mathcal{D}(\mathcal{N}_i^c)),
\end{equation}
where, $\mathcal{N}_i^c=\{j\in\mathcal{N}:\frac{\partial \mathbf{F}_i}{\partial x_j}=\mathbf{0}\}$ and $\Im(\cdot)$ denotes the image of the map $\mathcal{D}$.

\emph{Relaxation Process}: The relaxation process is an iterative procedure, with following steps
\begin{itemize}
\item Step 1: (Initialization of the relaxation process) Set $I=1$ and guess an initial waveform $\{\mathbf{y}_i^0(t): t\in [0\quad T]\}$ such that $\mathbf{y}^0_i(0)=\mathbf{y}_{0i}$, $i=1\cdots,m$.
\item Step 2: (Analyzing the decomposed system at the I-th WR iteration) For each $i=1,\cdots,m$, set
\begin{equation}\label{dk1}
\mathbf{d}_i^I(t)=\equiv((\mathbf{y}_{j_i}^{I-1})^T,\cdots,(\mathbf{y}_{j_{N_i}}^{I-1})^T)^T,
\end{equation}
and solve the subsystem
\begin{equation}\label{dk2}
\dot{\mathbf{y}}^I_i=\mathbf{F}_i(\mathbf{y}^I_i,\mathbf{d}_i^I(t),\Lambda_i,t),
\end{equation}
over the interval $[0,T_s]$ with initial condition $\mathbf{y}^I_i(0)=\mathbf{y}_{0i}$, to obtain $\{\mathbf{y}^I(t) : t \in [0,T_s]\}$.
\item Step 3 Set $I = I + 1$ and go to step 2 until satisfactory convergence is achieved..
\end{itemize}
The general conditions for convergence of WR for a system of differential algebraic equations (DAEs) can be found in \cite{waveform1,AWR}. Here, we recall a result from \cite{AWR} specializing it for a system
of differential equations.

\begin{proposition}Convergence of WR for ODE's (see \cite{waveform1} for proof):\label{WRConv} Given that the
system (\ref{complexsys}) is Lipschitz (condition~\ref{LipF}), then for any initial piecewise continuous
waveform $\{\mathbf{y}_i^0(t): t\in [0,T_s]\}$ such that $\mathbf{y}^0_i(0)=\mathbf{y}_{0i}$ (see definition (\ref{icfory})), $i=1\cdots,m$,
the WR algorithm converges to the solution of (\ref{complexsys}) with initial condition $\mathbf{x}_0$.
\end{proposition}

A more intuitive analysis of error at each waveform iteration is described in \cite{AWR}.
Let $ \bar{\mathbf{y}}$
be the exact solution of the differential equation (\ref{decomp}) and define $ E_I$ to be the error of the $ I $-th iterate, that is
\begin{equation}
    E_I(t)= \mathbf{y}^I(t) - \bar{\mathbf{y}}(t).
\end{equation}
As shown in \cite{AWR}, the error $|E_{I}| $ on the interval $ [0, T] $ is bounded as follows
\begin{equation} \label{BoundsTheo}
    |E_I(t)| \leq \frac{C^I \eta^I T^I}{I!} |E_0(t)|,
\end{equation}
with $ C = e^{\mu T}$. Here $C$ and $\eta$
are related to the Lipschitz constants of the waveform relaxation operator~\cite{AWR}. It is important to note here that the $I!$ in the denominator dominates the error. Thus, with enough iterations one can make the error fall below any desired threshold. It is also evident from Eq.~\ref{BoundsTheo} that the error of standard waveform relaxation crucially depends on $ T $. The longer the time interval, the greater is the number of iterations needed to bound the error below a desired tolerance.
Based on this observation, a novel ``adaptive'' version of waveform relaxation has been developed in~\cite{AWR}, which we refer to as adaptive waveform relaxation (AWR). The idea here is
to perform waveform relaxation in ``windows'' of time that are picked so as to reduce $I$ in Eq.~\ref{BoundsTheo}. Specifically, one can pick small time intervals for computation when the solution to (\ref{complexsys}) changes significantly (implying $E_{0}(t)$ is large) and pick large intervals when the solution changes little (consequently $E_{0}(t)$ is small). The solution from one time interval
is extrapolated to the next using a standard extrapolation formula~\cite{Stoer} and the initial error is estimated using,
\begin{equation} \label{E0form}
    \tilde{E}_{I+1, 0}(t) = \frac{\phi^{(l)}(x^{I+1}(\xi), x^{I+1}(\xi))}{(l+1)!} \, \omega(t),
\end{equation}
where,
\begin{equation} \label{omegaform}
    \omega(t) = (t - t_0)(t - t_1) \hdots (t - t_l).
\end{equation}
Here $t_{0}, t_{1},\hdots,t_{l}$ are points through which one passes the extrapolating polynomial~\cite{AWR}. Note that, $ \phi^{(l)} = \frac{d^l\phi}{dt^l}$
is the $ l $-th derivative of the waveform relaxation operator $ \phi $ with respect to $ t $ (see.~\cite{Stoer,AWR}).

The algorithm to compute the length of the time windows is as follows:

\emph{Adaptive Waveform Relaxation}: To compute the time interval for $ \Delta T_{I+1} $, execute the following steps:
\begin{enumerate}
\item Set $ \Delta T_{I+1} = 2 \Delta T_I $ and $ \delta = \frac{1}{20} \Delta T_I $.
\item Evaluate $ \tilde{E}_{I+1, 0}(T_I + \Delta T_{I+1}) $ using Eqn.~\ref{E0form} to estimate $ \norm{E_{I+1, 0}} $
and compute $ \norm{\hat{E}_{I+1, r}} $ with the aid of the following equation,
\begin{equation} \label{Test}
    \norm{\hat{E}_{I+1, r}} = \frac{ \left(e^{\mu \Delta T_{I+1}} \eta \Delta T_{I+1}\right)^r}{r!}
        \norm{E_{I+1, 0}}.
\end{equation}
\item If $ \norm{\hat{E}_{I+1, r}} > \varepsilon $ and $ \Delta T_{I+1} > \frac{1}{2} \Delta T_I $, set $ \Delta T_{I+1} = \Delta T_{I+1} - \delta $ and repeat step~2.
\end{enumerate}
We define the minimal window length to be $ \Delta T_{I+1} = \frac{1}{50} T $. The above procedure gives a sequence of time intervals $ [0, T_1] $, $ [T_1, T_2] $, $ \dots $, $ [T_{\nu-1}, T_{\nu}] $, where $ T_{\nu} = T $,  on which WR is performed (as described in relaxation process) with an initial ``guess'' waveform provided by an extrapolation of the solution on the previous interval ~\cite{AWR}. AWR is found to accelerate simulations by orders of magnitude over traditional WR ~\cite{AWR}. In this work, we propose to use AWR for simulating the set of differential equations that arise from intrusive polynomial chaos. As mentioned before, the curse of dimensionality gives rise to a combinatorial number of equations~\cite{gPC} making AWR particularly attractive.

While the convergence of WR or AWR is guaranteed  irrespective of how the system is decomposed in the assignment-partitioning step, the rate of convergence depends on the
decomposition~\cite{AWR}. For a given nonlinear system, determining a decomposition that leads to an optimal rate of AWR convergence is an NP-complete problem~\cite{AWR}. Ideally, to minimize the number of iterations required for convergence, one would like to place strongly interacting equations/variables on a single processor, with weak interactions between the variables or equations on different processors. We show in~\cite{AWR}, that spectral clustering~\cite{Tutorial} along with horizontal vertical decomposition~\cite{igorcdc} is a good heuristic for decomposing systems for fast convergence in WR and AWR. For this task, we now discuss a novel decentralized spectral clustering approach~\cite{ref:wave} that when coupled with AWR~\cite{AWR} provides a powerful tool for simulating large dynamical systems.

\subsection{Graph Decomposition}\label{specclus}
The problem of partitioning the system of equations (\ref{complexsys}) into subsystems based on how they interact or are coupled to each other, can be formulated as a graph decomposition problem. Given the set of states $x_1,\cdots,x_n$ and some
 notion of dependence $\overline{w}_{ij}\ge 0,i=1,\cdots,n,j=1,\cdots,n$ between pairs of states, an undirected graph $G=(V,E)$ can be constructed. The vertex set~$V =
\{1,\dots,n\}$ in this graph represent the states $x_i$, and the edge set is $E\subseteq V\times V$, where a weight~$\bar{w}_{ij} \geq 0$ is associated with each edge $(i,j)\in
E$, and $\overline{W}=[\bar{w}_{ij}]$ is the $n\times n$ weighted adjacency matrix
of~$\mathcal{G}$. In order to quantify coupling strength $\bar{w}_{ij}$ between nodes or states, we propose to use
\begin{equation}\label{similiarity}
\bar{w}_{ij}=\frac{1}{2}[|\overline{J}_{ij}|+|\overline{J}_{ji}|],
\end{equation}
where, $\overline{J}=[\frac{1}{T_s}\int_{t_0}^{t_0+T_s}J_{ij}(\mathbf{x}(t;\mathbf{\xi}_m),\mathbf{\xi}_m,t)dt]$,
is time average of the Jacobian,
\begin{equation}\label{Jac}
J(\mathbf{x},\mathbf{\xi},t)=\left[\frac{\partial f_i(\mathbf{x}(t;\mathbf{\xi}),\mathbf{\xi},t)}{\partial x_j}\right],
\end{equation}
computed along the solution $\mathbf{x}(t;\mathbf{\xi})$ of the system (\ref{complexsys}) for
nominal values of parameters $\mathbf{\xi}_m$. Use of system Jacobian  for
horizontal vertical graph decomposition can also be found in \cite{igorcdc}.

We will now discuss, spectral clustering (see~\cite{Tutorial} a popular graph decomposition/clustering approach that allows one to partition a undirected graph given its adjacency matrix $\overline{W}$ . In this method first a (normalized) graph Laplacian is constructed as follows ~\cite{Chung,Fiedler,Fiedler2},
\begin{align}
    L_{ij} = \begin{cases}
                1 & \mbox{if}\: i = j\\
                -\bar{w}_{ij}/\sum_{\ell=1}^N \bar{w}_{i\ell} & \mbox{if}\: (i,j) \in E\\
                0   & \mbox{otherwise}\,,
             \end{cases}
             \label{eq:ldef}
\end{align}
or equivalently as, $L= I-D^{-1}\overline{W}$ where $D$ is the diagonal matrix with the row sums of $\overline{W}$. The clustering assignment/decomposition is then obtained by computing the eigenvectors/eigenvalues of $L$. In particular, one uses the signs of the components of the second and higher eigenvectors to partition the nodes in the graph into clusters~\cite{Tutorial}. Traditionally, one can use standard matrix algorithms for eigenvectors/eigenvalues computation~\cite{GolubVanLoan96}. However, as the size of the  dynamical system or network (and thus corresponding adjacency matrix) increases, the execution of these standard algorithms becomes infeasible on monolithic computing devices. To address this issue, distributed eigenvalue/eigenvector computation methods have been developed, see for example \cite{KempeMcSherry08}.

In~\cite{ref:wave},  a wave equation based distributed algorithm to partition large graphs has been developed which computes the  partitions without constructing the entire adjacency matrix $\overline{W}$ of the graph~\cite{Tutorial}. In this method one ``hears'' clusters in the graph by computing the frequencies (using Fast Fourier Transform (FFT) locally at each node) at which the graph ``resonates''. In particular, one can show that these ``resonant frequencies'' are related to the eigenvalues of the graph Laplacian $L$ (\ref{eq:ldef}) and the coefficients of FFT expansion are the components of the eigenvectors. Infact, the algorithm is provably equivalent to the standard spectral clustering~\cite{Tutorial}, for details see ~\cite{ref:wave}.

The steps of the wave equation based clustering algorithm are as follows. One starts by writing the local update equation at node $i$ based on the discretized wave equation,
\begin{align}
    u_{i}(t) = 2u_{i}(t-1) - u_{i}(t-2) - c^2\displaystyle\sum_{j\in\mathcal{N}_i}L_{ij}
    u_{j}(t-1)\,
    \label{onenodewave}
\end{align}
where $u_{i}(t)$ is the value of $u$ at node $i$ at time $t$
and $L_{ij}$ are the local entries of the graph Laplacian.  At each node $i$, $u_{i}(0) = u_{i}(-1)$ is
set to a random number on the interval $\left[0,1\right]$. One
then updates the value of $u_{i}$ using Eqn. (\ref{onenodewave})
until $t=T_{max}$ (for a discussion on how to pick $T_{max}$
see~\cite{ref:wave}). Note that, $u_{i}(t)$ is a scalar
quantity and one only needs nearest neighbor information in
Eqn. (\ref{onenodewave}) to compute it. One then performs a local
FFT on $\left[u_{i}(1),\dots\dots,u_{i}(T_{max})\right]$ and
then assigns the coefficients of the peaks of the FFT to $v_{ij}$. Here
the frequency of the $j$-th peak is related to $\lambda_{j}$, the $j$-th eigenvalue
of the graph Laplacian $L$, and $v_{ij}$ is the $i$-th
component of the $j$-th eigenvector.

In~\cite{ref:wave}, it has been shown that for the wave speed $c<\sqrt{2}$ in
Eqn. (\ref{onenodewave}) above wave
equation based iterative algorithm is stable and converges. Moreover, the algorithm converges in $O(\sqrt{\tau})$ steps,
where $\tau$ is the mixing time of the Markov
Chain~\cite{ref:wave} associated with the graph $G$. The competing state-of-the-art
algorithm~\cite{KempeMcSherry08} converges in
$O(\tau(\log(n)^{2})$. For large graphs or datasets,
$O(\sqrt{\tau})$ convergence is shown to provide orders of
magnitude improvement over algorithms that converge in
$O(\tau(\log(n)^{2})$. For detailed analysis and
derivations related to the algorithm, we refer the reader
to~\cite{ref:wave}.

\section{Scalable Uncertainty Quantification Approach}\label{scaleUQ}
\begin{figure}[hbt]
\begin{center}
\includegraphics[scale=0.5]{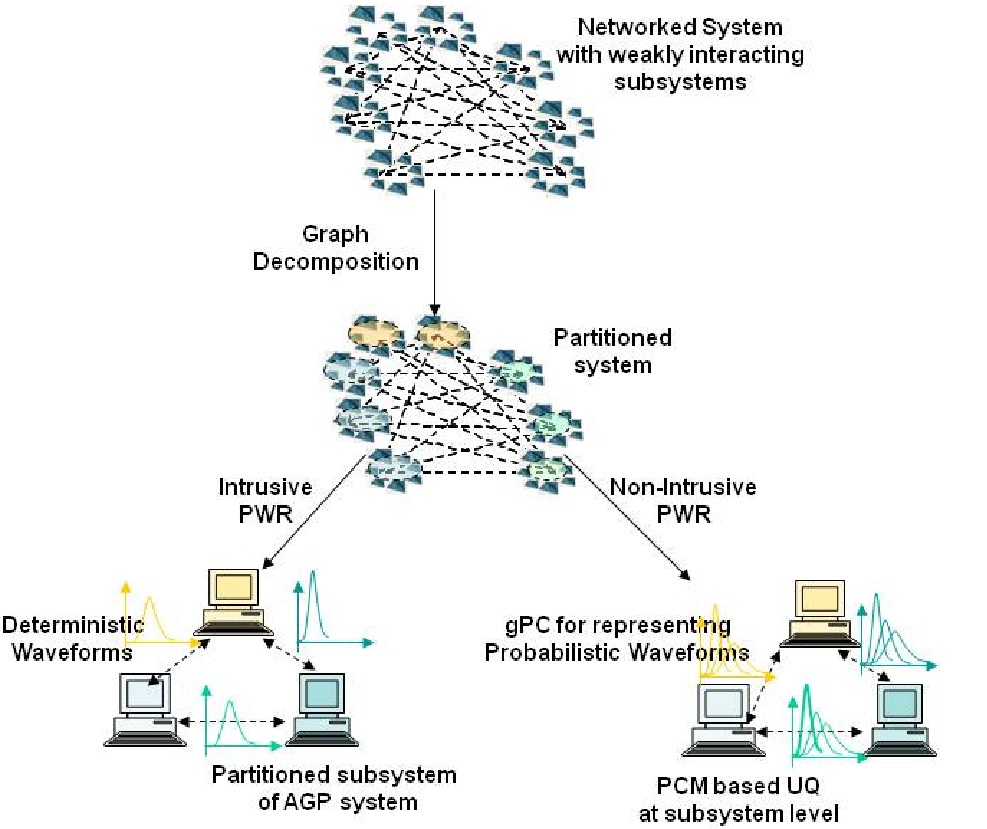}\\
\caption{Schematic of intrusive (left) and non-intrusive (right) PWR.}\label{schematicPWR}
\end{center}
\end{figure}
In this section we discuss how gPC and PCM can be integrated with WR scheme extending it to a
probabilistic setting. As mentioned earlier we refer to this iterative UQ approach as PWR.
Figure \ref{schematicPWR} shows the schematic of PWR framework. In the intrusive PWR, the subsystems obtained from
decomposing the original system are used to impose a decomposition on system obtained by Galerkin
projection based on the gPC expansion. Further the weak interactions are used to discard terms which are
expected to be  insignificant in the gPC expansion, leading to what we call an Approximate Galerkin Projected (AGP) system.
We then propose to apply standard or adaptive WR on the decomposed AGP system to accelerate the UQ computation.
In the non-intrusive form of PWR, rather than deriving the AGP system, one works directly with subsystems
obtained from decomposing the original system. At each waveform relaxation iteration we propose to
apply PCM at subsystem level, and use gPC to propagate the uncertainty among the subsystems. Note that unlike
intrusive PWR (where deterministic decoupling vectors or deterministic waveforms are exchanged),
in non-intrusive PWR, stochastic decoupling vector or probabilistic waveforms represented in gPC basis are
exchanged  between subsystems at each iteration.

We first describe the key technical ideas behind intrusive and non-intrusive
PWR though an illustration on a simple example in section \ref{sec:mainidea}.  These notions are fully
generalized later in sections  \ref{sec:GPdecomp}-\ref{waveprob}. We also prove the convergence
of PWR approach (in section \ref{sec:convPWR}), and in section \ref{sec:gainPWR} discuss the computational
gain it offers over standard application of gPC and PCM.

\subsection{Main Ideas of PWR}\label{sec:mainidea}
We illustrate the proposed PWR framework through an example of parametric
uncertainty in a simple system (\ref{complexsys}). Consider the following
coupled oscillator system:
\begin{eqnarray}
  \dot{x_1}&=&f_1(x_1,x_2,\omega_1,t)=\omega_1+K_{12} \sin(x_1-x_2),\notag\\
  \dot{x_2}&=&f_2(x_1,x_2,x_3,\omega_2,t)\notag\\
           &=& \omega_2+K_{21}\sin(x_1-x_2)+K_{23}\sin(x_3-x_2),\label{simpsys}\\
  \dot{x_3}&=&f_3(x_2,x_3,\omega_3,t)=\omega_1+K_{32} \sin(x_2-x_3)\notag
\end{eqnarray}
Here, $\omega_i$ is the uncertain angular frequency of the $i^{th}$ ($i=1,2,3$) oscillator.  We assume that parameters
$\omega_i$ are mutually independent, each having probability
density $\rho_i=\rho_i(\omega_i)$. The coupling matrix $K$ is
\begin{equation}\label{Kdef}
  K=\left(
      \begin{array}{ccc}
        0 & K_{12} & 0 \\
        K_{21} & 0 & K_{23} \\
        0 & 0 & K_{32} \\
      \end{array}
    \right)
\end{equation}
is assumed deterministic with $K_{ij}=\mathcal{O}(\epsilon),\epsilon \ll 1$, so that the three oscillators in (\ref{complexsys}) weakly interact with each other, i.e. the subsystem $2$ weakly affects subsystem $1$ and $3$, and vice versa.

\subsubsection{Approximate Galerkin projection for the simple example}\label{sec:agpsimpeg}
In standard gPC, states $x_i$ are expanded in a polynomial chaos basis as
\begin{equation}\label{expx1}
x_i^{\Sigma,P}(t,\mathbf{\omega}) =
	\sum_{j=1}^{N_{\Sigma}} a_{ji}(t)\Psi_j^{\Sigma,P}(\mathbf{\omega}),\quad i=1,2,3
\end{equation}
where, $\Psi_j^{\Sigma,P}\in W^{\Sigma,P}$, the $P$ variate polynomial chaos space formed over $\Sigma=\{\omega_1,\omega_2,\omega_3\}$  and $P=(P_1,P_2,P_3)$ determines the expansion order(see section \ref{gPC} for details). Note that in this expansion (\ref{expx1}), the system states are expanded in terms of all the random variables $\mathbf{\omega}$ affecting the entire system. From the structure of
system (\ref{complexsys}) it is clear that the $1^{st}$ subsystem is directly
affected by the parameter $\omega_1$ and indirectly by parameter $\omega_2$ through the
the state $x_2$. We neglect second order effect of $\omega_3$ on $x_1$.
A similar statement holds true for subsystem $3$, while subsystem $2$ will be
weakly influenced by $\omega_1$ and $\omega_3$ through states $x_1$ and $x_3$, respectively.
This structure can be used to simplify the Galerkin projection as follows. For $x_1$
we consider the gPC expansion over $\Sigma_1=\Lambda_1\bigcup\Lambda_1^c$,
\begin{equation}
	x_1^{\Sigma_1,P_1}(t,\omega_1,\omega_2) =
	\sum_{j=1}^{N_{\Sigma_1}} \overline{a}_{j1}(t)\Psi_j^{\Sigma_1,P_1}(\omega_1,\omega_2),
\end{equation}
where,
\begin{equation}
\Lambda_1=\{\omega_1\},\qquad \Lambda_1^c=\{\omega_2\}
\end{equation}
and $P_1=(P_{11},P_{12})$. Note that since $\omega_2$ weakly affects $x_1$, the order of expansion $P_{12}$ can be chosen to be smaller compared to $P_{11}$.
Similarly, one can consider simplification for $x_3^{\Sigma_3,P_3}(t,\omega_1,\omega_3)$. For $x_2$ following similar steps, we define
\begin{equation}
\Lambda_2=\{\omega_2\},\qquad \Lambda_2^c=\{\omega_1,\omega_3\}
\end{equation}
and $P_2=(P_{21},P_{22},P_{23})$. By similar argument, one will
choose $P_{21},P_{23}$ much smaller than $P_{22}$. We also introduce the following two projections associated with the state $x_2$:
\begin{equation}\label{proj11}
	\mathcal{P}^{2,i}(x_2^{\Sigma_2,P_2}) =
	\sum_{j=1}^{N_{\Sigma_i}} \left\langle x_2^{\Sigma_2,P_2},\Psi_{j}^{\Sigma_i,P_2} \right\rangle
	\Psi_{j}^{\Sigma_i,P_i}.
\end{equation}
where $i=1,3$  and $\langle \cdot,\cdot \rangle$ is the appropriate inner product on $W^{\Sigma,P}$ (see section \ref{approxPWR} for details).
With these expansions, and using standard Galerkin projection we obtain the following system of deterministic equations
\begin{equation}\label{agp1}
\dot{\overline{\mathbf{a}}}=\overline{\mathbf{F}}(\overline{\mathbf{a}},t),
\end{equation}
with appropriate initial conditions, where
\begin{eqnarray}\label{Fbar1}
	\overline{F}_{j1}(\overline{\mathbf{a}}) &=& \int_{\Gamma_{\Sigma}}f_1(x_1^{\Sigma_1,P_1},\mathcal{P}^{2,1}
	(x_2^{\Sigma_2,P_1}),\omega_1,t)\Psi_j^{\Sigma_1,P_1}(\mathbf{\omega})
	\mathbf{\rho}(\mathbf{\omega})d\mathbf{\omega},\notag, \\
\overline{F}_{j2}(\overline{\mathbf{a}})&=&\int_{\Gamma_{\Sigma}}f_2(x_1^{\Sigma_1,P_1},x_2^{\Sigma_2,P_2},x_3^{\Sigma_3,P_3},\omega_2,t)\Psi_j^{\Sigma_2,P_2}(\mathbf{\omega})
\mathbf{\rho}(\mathbf{\omega})d\mathbf{\omega},\notag\\
\overline{F}_{j3}(\overline{\mathbf{a}})&=&\int_{\Gamma_{\Sigma}}f_3(\mathcal{P}^{2,3}(x_2^{\Sigma_2,P_2}),x_1^{\Sigma_1,P_1},\omega_3,t)\Psi_j^{\Sigma_3,P_3}(\mathbf{\omega})
\mathbf{\rho}(\mathbf{\omega})d\mathbf{\omega},\notag
\end{eqnarray}
and $\overline{\mathbf{a}} =
(\overline{\mathbf{a}}_1, \overline{\mathbf{a}}_2, \overline{\mathbf{a}}_3)^T$,
with $\mathbf{a}_i = (\overline{a}_{i1},\cdots,\overline{a}_{iN_{\Sigma_{i}}})$ and
$\overline{\mathbf{F}} =
(\overline{\mathbf{F}}_1, \overline{\mathbf{F}}_2, \overline{\mathbf{F}}_3)^T$ with $\overline{\mathbf{F}}_i=(\overline{F}_{i1},\cdots,\overline{F}_{1N_{\Lambda_{i}}})$.
We will refer to (\ref{agp1}) as an approximate Galerkin projected (AGP) system.
The notion of AGP in more general setting is described in section \ref{approxPWR}.

\subsubsection{Intrusive PWR illustrated on the simple example}\label{sec:intpwrsimpeg}
In intrusive PWR, after performing the AGP explicitly,
the system (\ref{agp1}) is decomposed as
\begin{equation}\label{apsdecomp1}
\dot{\overline{a}}_{ij}=\overline{F}_{ji}(\overline{\mathbf{a}}_i,\overline{\mathbf{d}}_i,t),
\end{equation}
where $\overline{\mathbf{d}}_1 = \mathcal{P}^{2,1}(\overline{\mathbf{a}}_2)$,
$\overline{\mathbf{d}}_2 = (\overline{\mathbf{a}}_1, \overline{\mathbf{a}}_3)$
and $\overline{\mathbf{d}}_3=\mathcal{P}^{2,1}(\overline{\mathbf{a}}_2)$ are the decoupling
vectors (here we overloaded notation for $\mathcal{P}^{2,i}(\overline{\mathbf{a}}_2)$ to
imply the coefficients in expansion (\ref{proj11})). Note that the decomposition of
system (\ref{agp1}) is based on the decomposition of the original
system (\ref{simpsys}). Adaptive or standard WR described in section \ref{wave}, can
then be applied to solve the decomposed system (\ref{apsdecomp1}), iteratively. Since, the
the system (\ref{apsdecomp1}) is deterministic, deterministic waveforms or
deterministic decoupling vectors $\overline{\mathbf{d}}_i,i=1,2,3$
are exchanged in each WR iteration (see Figure \ref{schematicPWR} for an illustration).

\subsubsection{Non-Intrusive PWR illustrated on the simple example}\label{sec:nonintpwrsimpeg}
In the non-intrusive form of PWR, rather than deriving the AGP, one works directly with subsystems obtained from decomposing the original system. The main idea here is to apply PCM at subsystem level at each PWR iteration, use gPC to represent the probabilistic waveforms and iterate among subsystems using these waveforms. Recall that in standard PCM approach (\ref{PCM}), the coefficients $\overline{a}_{m}^i(t)$ are obtained by calculating
integral
\begin{eqnarray}
\overline{a}_{mi}(t)&=&\int x_{i}^{\Lambda_i,P_i}(t,\mathbf{\omega})\Psi_m^{\Lambda_i,P_i}(\mathbf{\omega})\mathbf{\rho}_{\Lambda_i}(\mathbf{\omega})d\mathbf{\omega}\label{z4approx1}
\end{eqnarray}
The integral above is typically calculated by using a quadrature formula and
repeatedly solving the $i^{th}$ subsystem over an appropriate collocation grid 
$\mathcal{C}^i(\Sigma_i)=\mathcal{C}^i(\Lambda_i)\times\mathcal{C}^i(\Lambda_i^c)$,
where, $\mathcal{C}^i(\Lambda_i)$
is the collocation grid corresponding to parameters $\Lambda_i$ (and let $l_s$
be the number of grid points for each random parameter in $\Lambda_i$),
$\mathcal{C}^i(\Lambda_i^c)$ is the collocation grid corresponding to
parameters $\Lambda^c_i$ (and let $l_c$ be the number of grid points for each
random parameter in  $\Lambda^c_i$ ). Since, the behavior of $i^{th}$ subsystem is
weakly affected by the parameters $\Lambda^c_i$, we can take a sparser grid
in $\Lambda^c_i$ dimension, i.e. $l_c<l_s$, as we took lower order expansion for these random variables
in section \ref{sec:agpsimpeg}. Below we outline key steps in non intrusive PWR:
\begin{itemize}
\item Step 1: (Initialization of the relaxation process with no coupling effect): Set $I=1$, guess an initial waveform $x_i^0(t)$ consistent with initial condition. Set
$\mathbf{d}_{1}^1=x^0_{2},\quad \mathbf{d}_{2}^1=(x^0_{1},x^0_3),\quad \mathbf{d}_{3}^1=x^0_2,$ and solve
\begin{equation}
  \dot{x}_i^1=f_i(x_i^1,\mathbf{d}_{i}^1(t),\omega_i,t),
\end{equation}
with an  initial condition $x^1_i(0)=x_i^0(0)$ on a collocation grid $\mathcal{C}^i(\Lambda_i)$. Determine the gPC expansion $x_{i}^{\Lambda_i,P_i,1}(t,\cdot)$ by computing the expansion coefficients from the quadrature formula (\ref{z4approx1}).
\item Step 2: (Initialization of the relaxation process, incorporating first level of coupling effect):
Set $I =2$ and let $\mathbf{d}_{1}^2=x_{2}^{\Lambda_2,P_2,1},\quad \mathbf{d}_{2}^2=(x_{1}^{\Lambda_1,P_1,1},x_{3}^{\Lambda_3,P_3,1}),\quad \mathbf{d}_{3}^2=x_{2}^{\Lambda_2,P_2,1}$ be
the \emph{stochastic decoupling vectors}. Solve
\begin{equation}
  \dot{x}_i^2=f_i(x_i^2,\mathbf{d}_{i}^2(t,\cdot),\omega_i,t),
\end{equation}
over a collocation grid $\mathcal{C}^i(\Sigma_i)$ to obtain
$x_i^{\Sigma_i,P_i,2}(t,\cdot)$. From now on we shall denote the solution
of the $i^{th}$ subsystem at $I^{th}$ iteration by  $x_{i}^{\Sigma_i,P_i,I}$.

\item Step 3 (Analyzing the decomposed system at the I-th iteration):
Set the decoupling vectors,
$\mathbf{d}_{1}^I=\mathcal{P}^{2,1}(x_{2}^{\Sigma_2,P_2,I-1}),\quad \mathbf{d}_{2}^I=(x_{1}^{\Sigma_1,I-1},x_{3}^{\Sigma_3,P_3,I-1})$, $\mathbf{d}_{3}^I=\mathcal{P}^{2,3}(x_{2}^{\Sigma_2,P_2,I-1})$ and solve
\begin{equation}
  \dot{x}_i^I=f_i(x_i^I,\mathbf{d}_{i}^I(t,\cdot),\omega_i,t),
\end{equation}
over a collocation grid $\mathcal{C}^i(\Sigma_i)$ and obtain the expansion $x_i^{\Sigma_i,P_i,I}(t,\cdot)$.

\item Step 4 (Iteration) Set $I = I + 1$ and go to step 5 until satisfactory convergence has been achieved.
\end{itemize}
Note that in above non-intrusive PWR, the decoupling vectors are stochastic and so at each iteration
\emph{probabilistic waveforms} are exchanged between subsystems ((see Figure \ref{schematicPWR} for an illustration).
We next generalize the intrusive and non-intrusive PWR introduced above, in the forthcoming sections.

\subsection{Decomposition of Galerkin Projected System}\label{sec:GPdecomp}
We begin by revisiting the complete Galerkin system (\ref{galproj}).
To apply WR, recall that the first step is the assignment-partitioning (see section \ref{wave}).
There are two possible approaches for partitioning the complete Galerkin system.
One can first split the original dynamical system (\ref{complexsys}), and then use this decomposition
to partition the complete Galerkin projection (\ref{galproj}) by assigning
the model coefficients in  (\ref{expx}) for each state to the cluster to which state is assigned to
while decomposing system (\ref{complexsys}). As previously explained in section \ref{specclus}, the
partitioning is performed by representing the dynamical system (\ref{complexsys}) as a graph with the
symmetrized time averaged Jacobian (\ref{similiarity}) as the weighted adjacency matrix.
One can then apply the wave equation based decentralized clustering algorithm
outlined in section \ref{specclus}.

Alternatively, one can perform this decomposition directly on the complete Galerkin projection (\ref{galproj}). Let the symmetrized time averaged Jacobian for the resulting system (\ref{galproj}) be,
\begin{equation}\label{similiarity_gpc}
\tilde{w}_{ij}=\frac{1}{2}[|\tilde{J}_{ij}|+|\tilde{J}_{ji}|],
\end{equation}
where, $\tilde{J}=[\frac{1}{T_s}\int_{t_0}^{t_0+T_s}J^{'}_{ij}(a(t),t)dt]$, is time average of the Jacobian,
\begin{equation}\label{Jac_gpc}
J^{'}(\mathbf{a},t)=\left[\frac{\partial F_{ik}(\mathbf{a}(t))}{\partial a_{ik}}\right],
\end{equation}
computed along the solution $\mathbf{a}(t)$ of the system (\ref{galproj}). This gives,
\begin{equation}\label{Jac_int}
J^{'}(\mathbf{a},t)= \int_{\Gamma_{\Sigma}}\frac{\partial f_k(\mathbf{x}^{\Sigma,P}(\mathbf{\xi},t),\mathbf{\xi}_k,t)}{\partial a_{jk}}\Psi_i^{\Sigma,P}(\mathbf{\xi})\mathbf{w}_{\Sigma}(\mathbf{\xi})d\mathbf{\xi}.
\end{equation}
Taylor expanding $f_k(\mathbf{x}(\mathbf{\xi},t),\mathbf{\xi}_k,t)$ locally, gives,
\begin{equation}\label{Jac_int}
J^{'}(\mathbf{a},t)\approx J(\mathbf{a},t).
\end{equation}
Thus, one expects to get similar results by performing clustering on the original system (in (\ref{complexsys})) to that obtained based on complete Galerkin system (\ref{galproj}). Since the dimensionality of system (\ref{complexsys}) is much lower than that of system (\ref{galproj}), the first decomposition is less computationally challenging than the latter. In this work, we will use the original system to determine the decomposition and use that to impose the partition of the Galerkin projection.
Given the decomposition of system (\ref{galproj}), one can use the WR or its adaptive form to
simulate the system in a parallel fashion. However, before doing this one can further exploit
the weak interaction between subsystems to reduce the dimensionality of the complete Galerkin system,
as described in section \ref{sec:agpsimpeg}. We next describe this approximate Galerkin projection in a
more general setting.

\subsection{Approximate Galerkin Projection and Intrusive Probabilistic Waveform Relaxation}\label{approxPWR}
Recall that in the gPC expansion (\ref{expx}), all the system states are expanded in terms of random variables affecting the entire system. However, the $i-$th subsystem in the decomposition (see Eq. (\ref{decomp})) is directly affected by the parameters $\Lambda_i$ (see definition (\ref{lamdef})) and indirectly by other parameters through the decoupling vector (see definition (\ref{ddef})). We shall denote by
\begin{equation}\label{complement}
\Lambda^c_i=\bigcup_{j\in\mathcal{N}_i}\Lambda_{j},
\end{equation}
the set of parameters which indirectly affect the $i-$th subsystem through the immediate neighbor interactions and by
\begin{equation}\label{complement}
\Sigma_i=\Lambda_i\cup\Lambda_i^c,
\end{equation}
the set of parameters that directly and indirectly (through nearest neighbor interaction) affect the $i-$th subsystem. Under the hypothesis that the $i-$th subsystem is dynamically weakly coupled with its nearest neighbors, uncertainty in parameters $\Lambda_c^i$ will weakly influence the states in $i-$th subsystem through the decoupling vector, while the uncertainty in parameters $\Sigma\setminus \Sigma_i$ can be neglected. To capture this effect, consider a $P$-variate space (analogous to the $P$-variate space introduced in the section \ref{gPC})
\begin{equation}\label{polyspacegensub}
W^{\Lambda,P}\equiv \bigotimes_{|\mathbf{d}|\in \mathbb{P}}W^{k_i,d_{k_i}},
\end{equation}
formed over any random parameter subset $\Lambda=\{\xi_{i_1},\xi_{i_2},\cdots,\xi_{i_n}\}\subset \Sigma$. We shall denote the basis elements of $W^{\Lambda,P}$ by $\Psi_{i}^{\Lambda,P},i=1,\cdots,N_{\Lambda}$, where $N_{\Lambda}=\mbox{dim}(W^{\Lambda,P})$.  Note that for any $\Lambda_1\subset\Lambda_2\subset \Sigma$,
\begin{equation}\label{sub}
W^{\emptyset}\subset W^{\Lambda_1,P_1}\subset W^{\Lambda_2,P_2}\subset W^{\Sigma,P},
\end{equation}
where, $W^{\emptyset}=\{0\}$ is the $P-$variate space corresponding to the empty set. Also, recall that $P_1,P_2$ are vectors which control the expansion order in gPC expansion. The inner product on $W^{\Sigma,P}$ induces an inner product on $W^{\Lambda,P}$ as follows
\begin{equation}\label{inducedip}
<X_1(\mathbf{\xi}),X_2(\mathbf{\xi})>_{\Lambda}=\int_{\Gamma_{\Lambda}}X_1(\mathbf{\xi}) X_2(\mathbf{\xi}) \mathbf{\rho}_{\Lambda}(\mathbf{\xi})d\mathbf{\xi},
\end{equation}
for any $X_1(\mathbf{\xi}),X_2(\mathbf{\xi})\in  W^{\Lambda,P}$.
Using this inner product, we introduce a projection operator
\begin{equation}\label{Projdef}
Pr_{\Lambda_1}^{\Lambda_2}:W^{\Lambda_2,P_2}\rightarrow W^{\Lambda_1,P_1},
\end{equation}
such that for any $X(\mathbf{\xi})\in W^{\Lambda_2,P_2}$
\begin{equation}\label{proj}
Pr_{\Lambda_1,P_1}^{\Lambda_2,P_2}(X)(\mathbf{\xi})=\sum_{i=1}^{N_{\Lambda_1}}<X,\Psi_{i}^{\Lambda_1,P_1S}>_{\Lambda}\Psi_{i}^{\Lambda_1,P_1}(\mathbf{\xi}).
\end{equation}
With respect to the given decomposition  $\mathcal{D}$ imposed on the system (see section \ref{wave}), we define a projection operator $\mathcal{P}^{i,j}$ indexed by subsystem $i$ and state $x_j$
\begin{equation}
\mathcal{P}^{i,j}\equiv \begin{cases}  \begin{array}{c}
                                Pr_{\Sigma_i,P_i}^{\Sigma,P},\quad \mbox{if} \quad \mathcal{D}(j)=i,\\
                                Pr_{\Lambda_{D(j)}\bigcup(\Lambda_{D(j)}^c\bigcap \Lambda_{i}),P_i}^{\Sigma,P}\quad \mbox{if} \quad \mathcal{D}(j)\neq i,\mathcal{N}_{i}\bigcap \mathcal{N}_{\mathcal{D}(j)}\neq\emptyset, \\
                                Pr_{\emptyset}^{\Sigma,P} \quad \mbox{if} \quad \mathcal{D}(j)\neq i, \mathcal{N}_{i}\bigcap \mathcal{N}_{\mathcal{D}(j)}=\emptyset.
                              \end{array}\notag
                              \end{cases}
\end{equation}

\begin{remark}\label{adaptiveexp} For any subsystem $i$, since the parameters $\Lambda_i^c$ weakly affect it, we can adaptively select component of
vector $P_i=(P_{i1},\cdots,P_{i,|\Sigma_i|})$ so that a lower order expansion is used in components corresponding to
random variables in $\Lambda_i^c$.
\end{remark}

For any subsystem $i$ and a vector $\mathbf{x}^{\Sigma,P}(\xi,t)=(x_{k_1}^{\Sigma,P}(\xi,t),\cdots,x_{k_n}^{\Sigma,P}(\xi,t))$ (where $x_{i}^{\Sigma,P}(\xi,t)$ is standard gPC expansion (\ref{expx})), we associate a vector valued projection operator as follows
\begin{equation}\label{vecP}
\mathcal{P}^{i}(\mathbf{x}^{\Sigma,P})=(\mathcal{P}^{i,k_1}(x_{k_1}^{\Sigma,P}),\cdots,\mathcal{P}^{i,k_n}(x_{k_n}^{\Sigma,P})).
\end{equation}
In terms of these operators, for any state $x_k$, an approximate Galerkin projected equation is defined as,
\begin{equation}\label{kth}
\frac{d}{dt}\mathcal{P}^{i,k}[x_k^{\Sigma,P}(\xi,t)]=f_k(\mathcal{P}^{i}(\mathbf{x}^{\Sigma,P}(\mathbf{\xi},t)),\mathbf{\xi}_k,t),
\end{equation}
where, $i=\mathcal{D}(k)$ is the index of the subsystem to which the state $k$ belongs. More, precisely the above system can be expressed as:
\begin{equation}\label{galproj1}
\dot{\overline{a}}^i_{jk}=\overline{F}_{jk}^i(\overline{\mathbf{a}},t),
\end{equation}
where, $\overline{\mathbf{a}}=(\overline{a}^{\mathcal{D}(1)}_{11},\cdots,\overline{a}^{\mathcal{D}(1)}_{N_{\Sigma_{\mathcal{D}(1)},1}},\cdots,\overline{a}^{\mathcal{D}(n)}_{1n},\cdots,\overline{a}^{\mathcal{D}(n)}_{N_{\Sigma_{\mathcal{D}(n)},n}})^T$
\begin{equation}\label{Fbar}
\overline{F}_{jk}^i=\int_{\Gamma_{\Sigma_i}}f_k(\mathcal{P}^{i}(\mathbf{x}^{\Sigma,P}(\mathbf{\xi},t)),\mathbf{\xi}_k,t)\Psi_j^{\Sigma_i,P_i}(\mathbf{\xi})
\mathbf{\rho}_{\Sigma_i}(\mathbf{\xi})d\mathbf{\xi},
\end{equation}
and, $j=1,\cdots,N_{\Sigma_i}$, $k=1,\cdots,n$ with  $i=\mathcal{D}(k)$. Let
\begin{equation}
\overline{\mathbf{F}}=(\overline{F}^{\mathcal{D}(1)}_{11},\cdots,\overline{F}^{\mathcal{D}(1)}_{N_{\Sigma_{\mathcal{D}(1)},1}},\cdots,\overline{F}^{\mathcal{D}(n)}_{1n},\cdots,\overline{F}^{\mathcal{D}(n)}_{N_{\Sigma_{\mathcal{D}(n)},n}})^T,\nonumber
\end{equation}

then the system (\ref{galproj1}) can be compactly written as
\begin{equation}\label{aps}
\dot{\overline{\mathbf{a}}}=\overline{\mathbf{F}}(\overline{\mathbf{a}},t),
\end{equation}
with appropriate initial condition (see expression \ref{galprojinit}) and will be referred to as the approximate Galerkin projected (AGP).
Using this generalization of AGP system, it is straightforward to generalize the intrusive PWR
introduced in section \ref{sec:intpwrsimpeg}.

\subsubsection{Intrusive PWR Algorithm}\label{compPWR}
In the intrusive PWR, one applies the WR to AGP system. As per discussion in section \ref{sec:GPdecomp},
the decomposition $\mathcal{D}$ on the original system (\ref{complexsys})  is used to imposes a decomposition
on the system (\ref{aps}) leading to
\begin{equation}\label{apsdecomp}
\dot{\overline{\mathbf{a}}_i}=\overline{\mathbf{F}}_i(\overline{\mathbf{a}}_i,\overline{\mathbf{d}}_i,t),
\end{equation}
for $i=1,\cdots,m$, where
\begin{equation}
\overline{\mathbf{a}}_i=(\overline{a}^{i}_{1k_1},\cdots,\overline{a}^{i}_{N_{\Sigma_{i},k_1}},\cdots,\overline{a}^{i}_{1k_{|\mathcal{C}_i|}},\cdots,\overline{a}^{i}_{N_{\Sigma_{i}},k_{|\mathcal{C}_i|}})^T,
\end{equation}
\begin{equation}
\overline{\mathbf{F}}_i=(\overline{F}^{i}_{1k_1},\cdots,\overline{F}^{i}_{N_{\Sigma_{i},k_1}},\cdots,\overline{F}^{i}_{1k_{|\mathcal{C}_i|}},\cdots,\overline{F}^{i}_{N_{\Sigma_{i}},k_{|\mathcal{C}_i|}})^T,
\end{equation}
$k_i\in\mathcal{C}_i$, and $\overline{\mathbf{d}}_i=(\overline{\mathbf{a}}_{j_i}^T,\cdots,\overline{\mathbf{a}}_{j_{N_i}}^T)^T$ is the decoupling vector (recall notation from section \ref{wave}).
One then follows the procedure for waveform relaxation or its adaptive version, as described in section \ref{wave}. Adaptive WR can lead to a significant increase in convergence of WR as demonstrated in \cite{AWR}, and would be illustrated later in the section \ref{examples}.

As discussed in section \ref{gPC}, the projection step (\ref{kth}) can  be very tedious and
in some cases not possible. Hence, applying waveform relaxation directly to the system (\ref{apsdecomp})
may not be practical in many instances. In the next section, we describe an alternative non-intrusive
approach using probabilistic collocation, which does not require the projection step (\ref{kth}) explicitly.

\subsection{Non-Intrusive Probabilistic Waveform Relaxation}\label{waveprob}
In terms of the projection operator (\ref{vecP}), we can rewrite each subsystem  in (\ref{decomp}) as
\begin{eqnarray}\label{decompparam}
  \dot{\mathbf{y}}_i &=&\mathbf{F}_i(\mathbf{y}_i,\mathcal{P}^i(\mathbf{d}_{i}(t,\cdot)),\Lambda_i,t),
\end{eqnarray}
where, $\mathbf{d}_{i}(t,\cdot)$ is the \emph{stochastic decoupling vector} or \emph{probabilistic waveform},
\begin{equation}\label{decoupstoc}
\mathbf{d}_{i}(t,\cdot)=((\mathbf{y}_{j_i}^{\Sigma_{j_1},P_{j_1}})^T,\cdots,(\mathbf{y}_{j_{N_i}}^{\Sigma_{j_{N_i}},P_{j_{N_i}}})^T)^T.
\end{equation}
where, we have explicitly indicated the dependence on the parameters ( see definition (\ref{ddef})).
Here for any $i=1,\cdots,m$, $\mathbf{y}^{\Sigma_i,P_i}=(x_{j^i_1}^{\Sigma_i,P_i},\cdots,x_{j^i_{M_i}}^{\Sigma_i,P_i})^T$, with
\begin{equation}\label{expnapprox}
x^{\Sigma_{i},P_i}_{j^i_k}(\mathbf{\xi},t)=\sum_{m=1}^{N_{\Sigma_i}} \overline{a}_{mj^i_k}(t) \Psi_m^{\Sigma_i,P_i}(\mathbf{\xi})=\mathcal{P}^{i,j^i_k}(x_{j^i_k}^{\Sigma,P}),
\end{equation}
By definition the coefficients $\overline{a}_{mj^i_k}(t)$ in above expansion satisfy the system (\ref{galproj1}). These coefficients can be obtained by using the quadrature formula (\ref{z4approx}), by repeatedly solving the system (\ref{decompparam}) over an appropriate collocation grid $C(\mathbf{l},n_i)$
\begin{equation}\label{fullgirdgen}
\mathcal{C}(\mathbf{l},n_i+n_i^c)=\mathcal{C}(\mathbf{o},n_i)\times\mathcal{C}(\mathbf{m},n_i^c),
\end{equation}
where, $\mathbf{l}=(\mathbf{o},\mathbf{m})$, $\mathcal{C}(\mathbf{o},n_i)=C_{o_1}^1\times\cdots \times C_{o_{n_i}}^1,$
is the collocation grid corresponding to parameters $\Lambda_i$, with $n_i=|\Lambda_i|$, $\mathbf{o}=(o_1,\cdots,o_{n_i})$, and $\mathcal{C}(\mathbf{m},n_i^c)=C_{m_1}^1\times\cdots \times C_{m_{n_i^c}}^1,$
is the collocation grid corresponding to parameters $\Lambda_i^c$, with $n_i^c=|\Lambda_i^c|$ and $\mathbf{m}=(m_1,\cdots,m_{n_i^c})$.
For simplicity  we take $o_{1}=\cdots=o_{n_i}=l_s$ and $m_{1}=\cdots=m_{n_i^c}=l_c$ for $i=1\cdots,m$.
Since, the behavior of $i-$th subsystem is weakly affected by the parameters $\Lambda^c_i$ through the decoupling vector, then consistent with
remark (\ref{adaptiveexp}) we can take
\begin{equation}\label{maxcond}
l_c<l_s,
\end{equation}
leading to an adaptive collocation grid for each subsystem. With this, we are ready to generalize the
non-intrusive PWR approach introduced in section \ref{sec:nonintpwrsimpeg}.

\subsubsection{Non-Intrusive PWR Algorithm}\label{PWR}
\begin{itemize}
   \item Step 1: Apply graph decomposition (see section \ref{nd} for details) to identify weakly interacting subsystems in the system (\ref{complexsys}).
\item Step 2 (Assignment-partitioning process): Partition (\ref{complexsys}) into $m$ subsystems (obtained in Step I) leading to system of equations given by (\ref{decomp}). Obtain, $\Lambda_i$, $\Lambda_i^c$ and $\Sigma_i$ for each subsystem, $i=1,\cdots,m$.
Choose the parameters $l_{si},l_{ci},P_i$.
\item Step 3: (Initialization of the relaxation process with no coupling effect): Set $I=1$ and , guess an initial waveform $\{\mathbf{y}_i^0(t): t\in [0,T_s]\}$ for each $i=1,\cdots,m$ consistent with initial condition (see Step 1 in relaxation process described in section \ref{wave}). Set
\begin{equation}\label{ydk0}
\mathbf{d}_{i}^1(t)=(\mathbf{y}_{j_1}(t),\cdots,\mathbf{y}_{j_{N_i}}(t)),
\end{equation}
$i=1,\cdots,m$, and solve for $\{\mathbf{y}_i^{\Lambda_i,P_i,1}(t), t \in [0,T_s]\}$ using
\begin{equation}
  \dot{\mathbf{y}}_i^1=\mathbf{F}^i(\mathbf{y}_i^1,\mathbf{d}_{i}^1(t),\Lambda_i,t),
\end{equation}
with an  initial condition $\mathbf{y}^1_i(0)=\mathbf{y}_i^0(0)$ on a collocation grid $C(\mathbf{o},n_i)$. Determine the gPC expansion $\mathbf{y}_{i}^{\Lambda_i,P_i,1}(t,\cdot)$ over $P-$variate polynomial space $W^{\Lambda_i,P_i}$ by computing the expansion coefficients from the quadrature formula (\ref{z4approx}).
\item Step 4: (Initialization of the relaxation process, incorporating first level of coupling effect): Set $I =2$ and for each $i=1,\cdots,m$, set
\begin{equation}\label{ydk1}
\mathbf{d}_{i}^2(t,\cdot)=(\mathbf{y}_{j_1}^{\Lambda_{j_1}, P_{j_1},1}(t,\cdot),\cdots,\mathbf{y}_{j_{N_i}}^{\Lambda_{j_{N_i}},P_{j_{N_i}},1}(t,\cdot)),
\end{equation}
and solve for $\{\mathbf{y}_i^{2}(t), t \in [0,T_s]\}$ from
\begin{equation}
  \dot{\mathbf{y}}_i^2=\mathbf{F}^i(\mathbf{y}_i^2,\mathbf{d}_{i}^2(t,\cdot),\Lambda_i,t),
\end{equation}
with an  initial condition $\mathbf{y}^2_i(0)=\mathbf{y}^0_i(0)$, over a collocation grid $C(\mathbf{l},n_i+n_i^c)$. Obtain the expansion $\mathbf{y}_i^{\Sigma_i,P_i,2}(t,\cdot)$ using (\ref{expnapprox}). From now on we shall denote the solution vector of the $i-$th subsystem at $I-$th iteration by  $\mathbf{y}_{i}^{\Sigma_i,P_i,I}$.

\item Step 5 (Analyzing the decomposed system at the I-th iteration):
For each $i=1,\cdots,m$,
set
\begin{equation}\label{ydk2}
\mathbf{d}_{i}^I=(\mathbf{y}_{j_1}^{\Sigma_{j_1}P_{j_1},(I-1)},\cdots,\mathbf{y}_{j_{N_i}}^{\Sigma_{j_{N_i}},P_{j_{N_i}},(I-1)}),
\end{equation}
and solve for $\{\mathbf{y}^I(t) : t \in [0,T_s]\}$ from
\begin{equation}
  \dot{\mathbf{y}}_i^I=\mathbf{F}^i(\mathbf{y}_i^I,\mathcal{P}^{i}(\mathbf{d}_{i}^I(t,\cdot)),\Lambda_i,t),
\end{equation}
with initial condition $\mathbf{y}^I_i(0)=\mathbf{y}^0_i(0)$ over a collocation grid $C(\mathbf{l},n_i+n_i^c)$. Obtain the expansion $\mathbf{y}_i^{\Sigma_i,P_i,I}(t,\cdot)$ using the expansions (\ref{expnapprox}).

\item Step 6 (Iteration) Set $I = I + 1$ and go to step 5 until satisfactory convergence has been achieved.
\end{itemize}

\subsection{Convergence of PWR} \label{sec:convPWR}
Below we prove that the iterative PWR approach converges. The proof is based on showing that the AGP system is Lipschitz if the orginal systems is  Lipschitz (see condition \ref{LipF}),  and then invoking standard WR convergence result (\ref{WRConv}).
\begin{proposition}Convergence of PWR: The intrusive and non-intrusive PWR algorithms described in sections \ref{compPWR} and \ref{PWR}, respectively converge.
\end{proposition}
\emph{Proof:} We prove the result for intrusive PWR. By construction, since non-intrusive PWR algorithm solves the AGP system (\ref{aps}) in a different way, the convergence result holds for non-intrusive PWR as well. Consider the AGP system(\ref{aps}) and let
\begin{equation}
\mathbf{a}_1=(\overline{a}^{1\mathcal{D}(1)}_{11},\cdots,\overline{a}^{1\mathcal{D}(1)}_{N_{\Sigma_{\mathcal{D}(1)},1}},\cdots,\overline{a}^{1\mathcal{D}(n)}_{1n},\cdots,\overline{a}^{1\mathcal{D}(n)}_{N_{\Sigma_{\mathcal{D}(n)},n}})^T,
\end{equation}
and
\begin{equation}
\mathbf{a}_2=(\overline{a}^{2\mathcal{D}(1)}_{11},\cdots,\overline{a}^{2\mathcal{D}(1)}_{N_{\Sigma_{\mathcal{D}(1)},1}},\cdots,\overline{a}^{2\mathcal{D}(n)}_{1n},\cdots,\overline{a}^{2\mathcal{D}(n)}_{N_{\Sigma_{\mathcal{D}(n)},n}})^T.
\end{equation}
Let for a given $k=1,\cdots,n$, $i=\mathcal{D}(k)$, then
\begin{equation}
\mathcal{P}^{i,k}(x_k^{l,\Sigma,P}(t,\mathbf{\xi}))=\sum_{j=1}^{N_{\Sigma_i}} \overline{a}_{jk}^{li}(t)\Psi_j^{\Sigma_i,P}(\mathbf{\xi}),
\end{equation}
and $\mathcal{P}^{i}(\mathbf{x}^{l,\Sigma,P})=(\mathcal{P}^{i,1}(x_1^{l,\Sigma,P}),\cdots,\mathcal{P}^{i,n}(x_n^{l,\Sigma,P}))$,
for $l=1,2$ and for simplifying notation we have dropped subscripts on $P$ vectors.  Then for each $k=1,\cdots,n$,$i=\mathcal{D}(k)$, $j=1,\cdots,N_{\Sigma_{i}}$,
\begin{eqnarray}
&&||\overline{F}^i_{jk}(\mathbf{a}_2)-\overline{F}^i_{jk}(\mathbf{a}_1)||\notag\\
&=&|\int_{\Gamma_{\Sigma}}(f_k(\mathcal{P}^{i}(\mathbf{x}^{\Sigma,P,2}),\xi,t)-f_k(\mathcal{P}^{i}(\mathbf{x}^{\Sigma,P,1}),\xi,t))\notag\\
&&\times \Psi_j^{\Sigma_i,P}(\mathbf{\xi})w_{\Sigma}(\mathbf{\xi})d\mathbf{\xi}|\notag\\
&\leq&\int_{\Gamma_{\Sigma}}L(\xi)(\sum_{m=1}^n\sum_{p=1}^{N_{\Sigma_{\mathcal{D}(m)}}}|(a_{pm}^{l,\mathcal{D}(m)}-a_{pm}^{2,\mathcal{D}(m)})\Psi_p^{\Sigma_{\mathcal{D}(m)},P}|)\notag\\
&&\times|\Psi_j^{\Sigma_i,P}(\mathbf{\xi})|w_{\Sigma}(\mathbf{\xi})d\mathbf{\xi}\notag\\
&\leq&\sum_{m=1}^n\sum_{p=1}^{N_{\Sigma_{\mathcal{D}(m)}}}L_{pj}^{i\mathcal{D}(m)}|(a_{pm}^{1,\mathcal{D}(m)}-a_{pm}^{2,\mathcal{D}(m)})|
\end{eqnarray}
where,
\begin{eqnarray}
L_{pj}^{iq}&=& \int_{\Gamma_{\Sigma}}L(\xi)|\Psi_p^{\Sigma_{q},P}(\xi)||\Psi_j^{\Sigma_i,P}(\xi)|w_{\Sigma}(\mathbf{\xi})d\mathbf{\xi}.
\end{eqnarray}
For a given $i=1,\cdots,n$ and $j=1\cdots,N_{\Sigma_{\mathcal{D}(i)}}$,  let
\begin{equation} L^g_{ij}=[L_{j1}^{\mathcal{D}(i)\mathcal{D}(1)}  \cdots  L_{jN_{\Sigma_{\mathcal{D}(1)}}}^{\mathcal{D}(i)\mathcal{D}(1)}  \cdots  L_{j1}^{\mathcal{D}(i)\mathcal{D}(n)}  \cdots  L_{jN_{\Sigma_{\mathcal{D}(n)}}}^{\mathcal{D}(i)\mathcal{D}(n)}],\nonumber
\end{equation}
 and define
\begin{equation}
\quad L^g_i=\left[
                                          \begin{array}{c}
                                          L_{i1}^g \\
                                            \vdots \\
                                           L_{iN_{\Sigma_{\mathcal{D}(i)}}}^g\\
                                          \end{array}
                                        \right], L^g=\left[\begin{array}{c}
           L^g_1\\
           \vdots \\
           L^g_n
         \end{array}\right],
\end{equation}
then,
\begin{eqnarray}\label{LipH}
||\overline{\mathbf{F}}(\mathbf{a}_2)-\overline{\mathbf{F}}(\mathbf{a}_1)||&\leq&\overline{L}||\mathbf{a}_2-\mathbf{a}_1||,
\end{eqnarray}
where, $\overline{L}=||L^g||$ is a matrix norm of $L^g$. Hence, the system (\ref{aps}) is Liptchitz.
Thus, given the original system (\ref{complexsys}) is Lipschitz (condition (\ref{LipF})), the approximate system (\ref{aps}) is also Lipschitz as shown above (\ref{LipH}). Hence, by proposition \ref{WRConv} (adaptation of Theorem 5.3 in (\cite{waveform1})), we conclude that PWR converges.

The question of how the decomposition of the system and the choice of the PWR algorithm parameters $P,l_s,l_c$ influence: 1) the rate of convergence of PWR, and 2) the approximation error (due to the truncation introduced in the AGP system (\ref{aps}), the use of adaptive collocation grid i.e. condition \ref{maxcond} and computation of the modal coefficients by the quadrature formula), needs to be further investigated.

\subsection{Scalability of PWR}\label{sec:gainPWR}
The scalability of non-intrusive PWR relative to full grid PCM is shown in Figure \ref{fig2}, where the ratio $\mathcal{R}_F/\mathcal{R}_I$ indicates the computation gain over standard full grid approach applied to the system (\ref{complexsys}) as a whole. Here $\mathcal{R}_F=l^{p}$ is the number of deterministic runs of the complete system (\ref{complexsys}), which comprises of $m$ subsystems each with $p_i,i=1,\cdots,m$ uncertain parameters, such that $p=\sum_{i=1}^mp_i$ and $l$ denotes the level of full grid. Similarly, $\mathcal{R}_I=1+\sum_{i=1}^ml_s^{p_i}+I_{\mbox{max}}(\sum_{i=1}^m l_s^{p_i}\bigotimes_{j\neq i}l_c^{p_j})$ is the total computational effort with PWR algorithm, where  $I_{\max}$ is the number of PWR iterations. Clearly, the advantage of PWR becomes evident as the number of subsystems $m$ and parameters in the network increases. Moreover PWR is inherently parallelizable.

\begin{figure}[hbt]
\begin{center}
\includegraphics[scale=0.3]{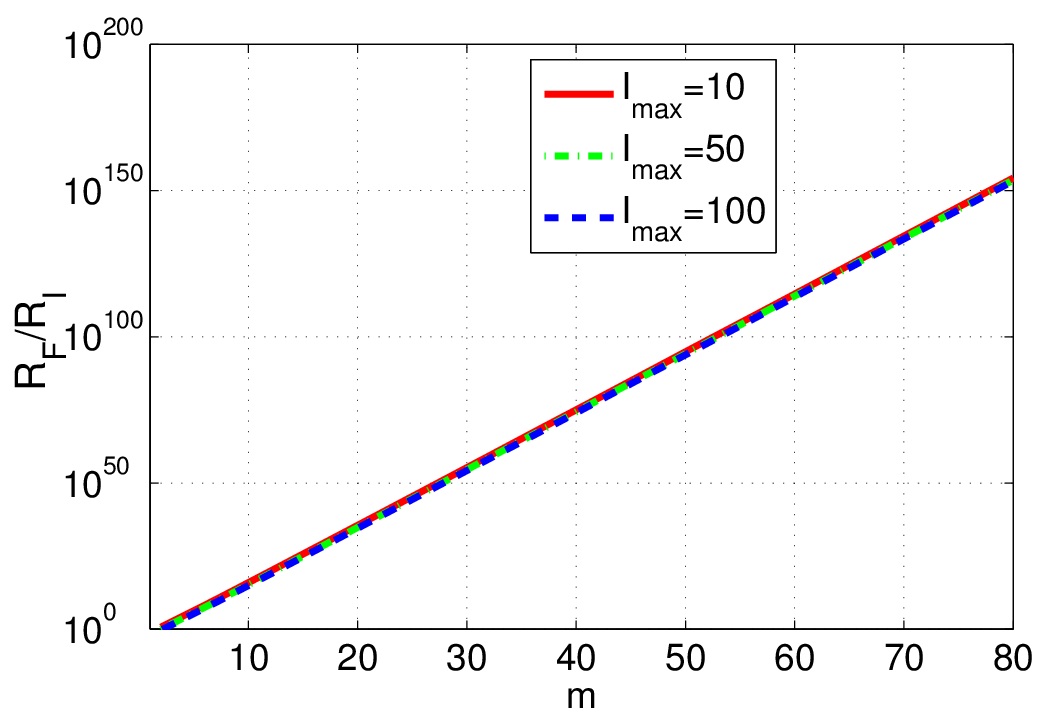}\\
\caption{Scalability of PWR algorithm, when implemented with full grid collocation as the subsystem level UQ method, with $p_i=5,\forall i=1,\cdots,m$, $l=l_s=5$, $l_c=3$ and $I_{\max}=10,50,100$. The computational gain of PWR becomes insensitive to $I_{\max}$, as the number of subsystems $m$ increase. }\label{fig2}
\end{center}
\end{figure}

\section{Example Problems}\label{examples}
In this section we illustrate intrusive and non-intrusive PWR through several examples of linear and nonlinear networked systems with increasing number of uncertain parameters. While most examples are of ODE's, we also give an example application of PWR to an algebraic system. This illustrates how in principle one can extend application of PWR to DAEs, just like WR approach extends to DAEs \cite{wave}). Through some examples we study how the strength of interaction between subsystems affects the convergence rate and the approximation error of PWR. In one of the examples related to building model, we also show how time-varying uncertainty can be incorporated into standard UQ framework by using Karhunen Loeve expansion. In all the examples, we compare solution accuracy of PWR with other UQ approaches (e.g. Monte Carlo and Quasi Monte Carlo methods), and wherever appropriate mention computation gain offered by PWR over the standard application of gPC and PCM.

\subsection{Stability Problem}
\begin{figure}[hbt]
\begin{center}
\includegraphics[width=5.7cm]{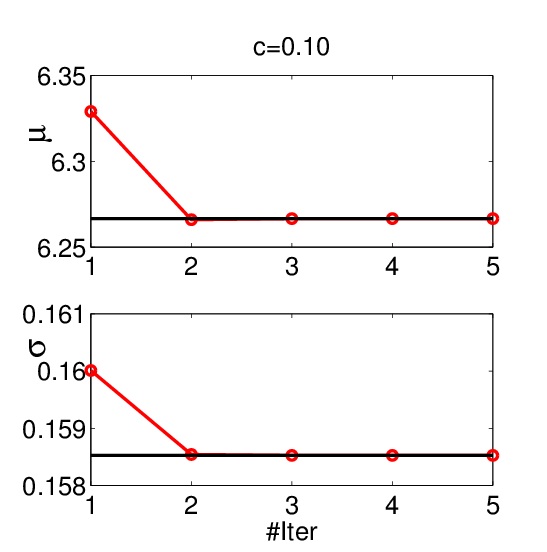}
\includegraphics[scale=0.33]{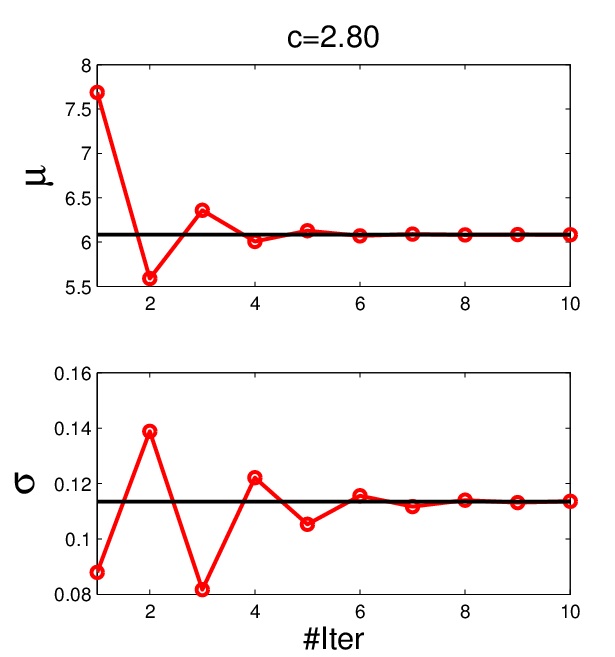}
\caption{Left Panel: Convergence of mean ($\mu$) and variance ($\sigma$) of $\lambda_{m}$ for $c=0.1$. Right panel: Convergence of mean and variance of $\lambda_{m}$ for $c=2.8$. Black line indicates the \emph{true} values.}\label{case4c}
\end{center}
\end{figure}

 We first consider a simple system, with two states $(x_1,x_2)\in \mathbb{R}^2$,
\begin{eqnarray}\label{simp}
\dot{x}_1&=&ax_1^2+cx_2^2-v_1,  \\
\dot{x}_2&=&cx_1^2+bx_2^2-v_2,
\end{eqnarray}
where, $c,v_1,v_2$ are fixed parameters, and $a,b$ are uncertain with Gaussian distribution $\mathcal{G}$ and tolerance $20\%$ (i.e.  $\sigma=0.2\mu$, where $\mu$ is the mean and $\sigma$ is the standard deviation of $\mathcal{G}$). The parameter $c$ determines the coupling strength between two subsystems described by the two equations. The output of interest is the stability of the system, which is determined by $\lambda_{m}$, the maximum eigenvalue of the Jacobian $ J(x_{10},x_{20};a,b,c)=\left(
     \begin{array}{cc}
       2ax_{10} & 2cx_{20} \\
       2cx_{10} & 2bx_{20} \\
     \end{array}
   \right)$,
where, $x_{10},x_{20}$ is the equilibrium point satisfying
\begin{eqnarray}
ax_{10}^2+cx_{20}^2-v_1&=&0, \notag \\
cx_{10}^2+bx_{20}^2-v_2&=&0.\label{simpeq}
\end{eqnarray}
Figure \ref{case4c} shows the UQ results obtained by applying PWR (with $l_s=5$, $l_c=3$ and $P_1=(5,3),P_2=(3,5)$) to iteratively solve the algebraic system (\ref{simpeq}). We make comparison with the \emph{true} (to imply more accurate result obtained by solving the complete system (\ref{simpeq})) distribution  of $\lambda_{m}$ obtained by using a full collocation grid on the parameter space $(a,b)$ with $l_a=5,l_b=5,P=(5,5)$. PWR converges to the true mean and variance as shown in the left and right panel of the Figure \ref{case4c} for two different values of $c$.  As the coupling strength $c$ increases (see right panel in Figure \ref{case4c}), the number of iterations required for the convergence increases, as expected.

\subsection{Building Example}
For energy consumption computation, a building can be represented in terms of a reduced order thermal network model of the form \cite{zheng2009},
\begin{equation}\label{Buildmodel}
\frac{d \mathbf{T}}{dt}=A(\mathbf{u}(t);\xi)T+B\left(
                       \begin{array}{c}
                         \mathbf{Q}_{e}(t) \\
                         Q_{i}(t) \\
                       \end{array}
                     \right)
\end{equation}
where, $\mathbf{T}\in R^n$ is a vector comprising of internal zone air temperatures, and internal and external wall temperatures; $A(\mathbf{u}(t);\xi)$ is the time dependent matrix with $\xi$ being parameters, $\mathbf{u}(t)$ is control input vector (comprising of zone supply flow rate and supply temperature), and vectors $\mathbf{Q}_e=(T_{amb}(t),Q_{s}(t))^T$  represent the external (outside air temperature and solar radiation), and $Q_i$ is the internal (occupant) load disturbances. We consider the problem of computing uncertainty in building energy consumption due to uncertainty in building thermal related properties and uncertain disturbance loads. These uncertainties can be categorized into: (i) static parametric uncertainty which include parameters such as wall thermal conductivity and thermal capacitance, heat transfer coefficient, window thermal resistance etc.; and (ii) time varying uncertainties which include the external and internal load disturbances.

Recall, that the traditional UQ approaches and PWR which builds on them, can only deal parametric uncertainty. To account for time varying uncertain processes, we employ Karhunen Loeve (KL) expansion \cite{KL}. The KL expansion allows representation of second order stochastic processes as a sum of random variables. In this manner, both parametric and time varying uncertainties can be treated in terms of random variables. We next demonstrate both intrusive and non-intrusive PWR methods.

\subsubsection{Two Zone Example}
\begin{figure}[htb]
\begin{center}
\includegraphics[scale=0.4]{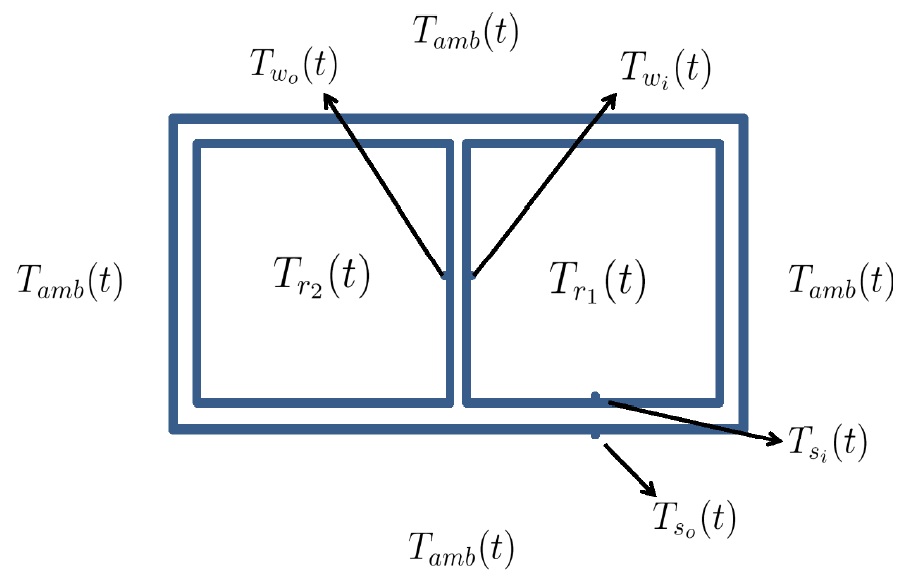}
\end{center}
\caption{Diagram of the two zone thermal model of a building. $T_{amb}(t)=293$K in this example.}
\label{Fig:two_zone}
\end{figure}
We first consider a simplified two zone building model as shown in Fig.~\ref{Fig:two_zone}. Here the state $\mathbf{T}$ is a $10$ dimensional vector comprising of internal wall temperatures and the internal zone air temperatures, where we have assumed that the outer wall surfaces are held at ambient temperature. We also assume that the ambient temperature and solar load are deterministic fixed quantities and there is no internal occupant load. Thus, in computing the uncertainty in the zone temperatures,  we only consider parametric uncertainty. Specifically, we assume that the heat transfer coefficient and the thermal conductivity of the walls in each zone have standard deviations of $10\%$ around their nominal values of $3.16 W/m^{2}/K$ and $4.65 W/m/K$, respectively. Thus, locally each zone is affected by two uncertain parameters, with heat transfer coefficient being a common (i.e. same) parameter and thermal conductivity being the other. Using complete Galerkin projection with $P_i=(2,2,2),i=1,2$, gives rise to a $60$ state ODE model. To apply WR/AWR to this system, we first identify the weakly interacting states. By construction the two zones weakly affect each other, which is identified by the spectral clustering~\cite{Tutorial} (or wave equation based clustering~\cite{ref:wave}) applied to the system (\ref{Buildmodel}). This decomposition is imposed on the complete Galerkin system, as explained in section \ref{compPWR}. As expected, we found that if ones applies spectral clustering to the complete Galerkin system instead, one recovers same decomposition.

\begin{figure}[htb]
    \begin{center}
   \subfigure[]{\includegraphics[scale=0.35]{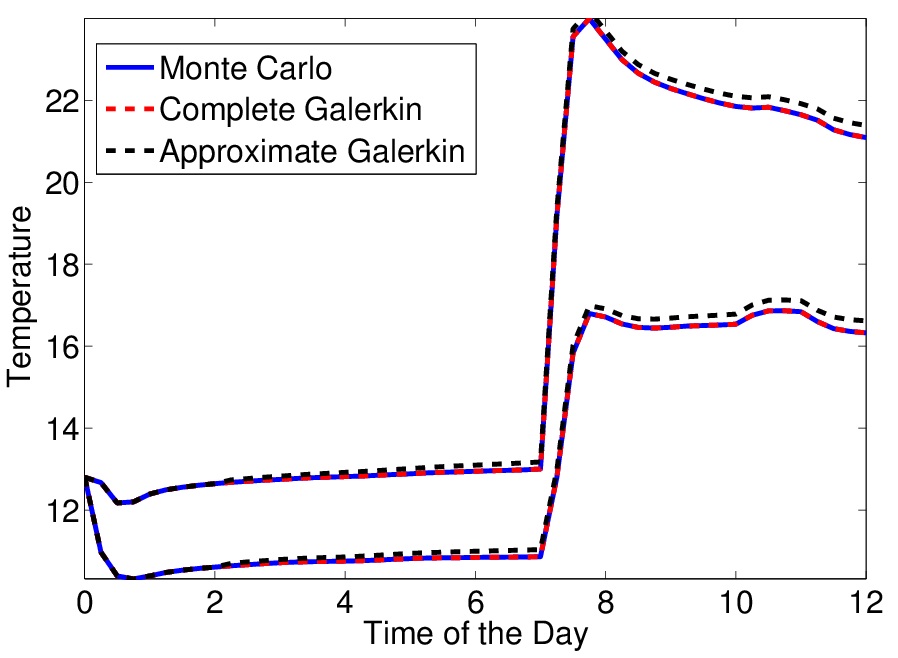}}
   \subfigure[]{\includegraphics[scale=0.5]{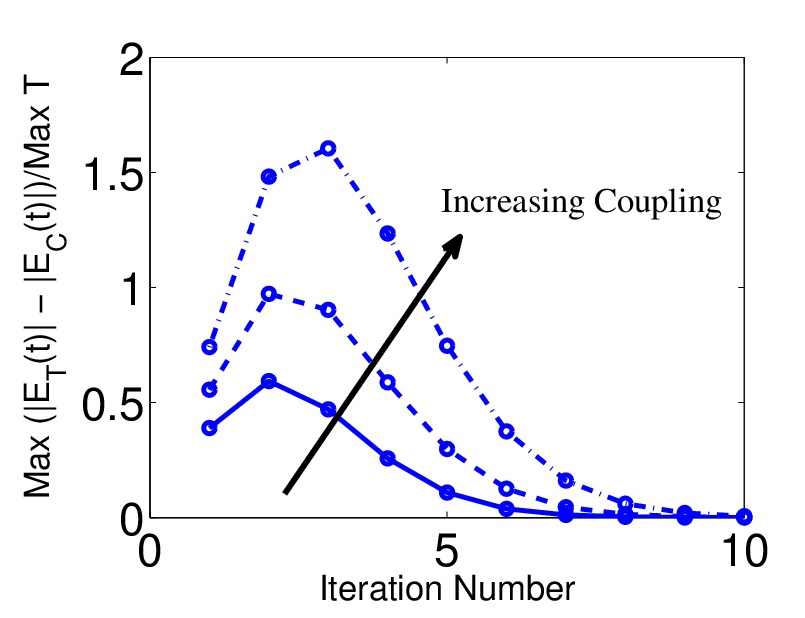}}
    \end{center}
    \caption{(a) Comparison of Monte Carlo, complete Galerkin projection and approximate Galerkin projection. (b)Normalized error in waveform relaxation as a function of iteration count with increasing coupling. Complete Galerkin and approximate Galerkin are shown. Approximate Galerkin system is found to have greater error as a function of iteration number.} \label{Fig:Completevstrunc}
\end{figure}
Treating $1000$ Monte Carlo samples as the truth, we compare the results of simulated full Galerkin projected system using both standard waveform relaxation~\cite{waveform1} as well as adaptive waveform relaxation~\cite{AWR} in Fig.~\ref{Fig:Completevstrunc}a). AWR provides a speed-up by a factor of $\approx 12$. In Fig.~\ref{Fig:Completevstrunc}a), one can visually see that the complete Galerkin Projection predicts the same temperature variation over $8$ hours as Monte Carlo based methods.

As explained before, one can further exploit the weak interaction between the two zones to reduce the overall number of equations in Galerkin projection. To construct the AGP system, we reduce the order of expansion for the random parameters indirectly affecting each zone so that $P_1=(2,2,1)$ and $P_2=(1,2,2)$. With this the number of equations in Galerkin projection reduces from $60$ to $50$. The resulting solution is shown in Fig.~\ref{Fig:Completevstrunc}a). We see that the error starts to grow as time increases. However, over $8$ hours the max error in the room temperatures is $5\times10^{-2}K$. Thus, despite reducing the computational effort, one can still get a fairly accurate answer.

Figure ~\ref{Fig:Completevstrunc}b) shows the effect of coupling (which is the reciprocal of the coefficient of thermal conductivity of the internal wall) on errors introduced in complete and approximate Galerkin projections. As expected, the approximate Galerkin projection has higher error (given by $E_{T}(t)$) than complete Galerkin projection (given by $E_{C}(t)$). Moreover this error is more pronounced at low iteration numbers. From the figure, it also clear that as the coupling increases, the number of iterations required for obtaining same solution accuracy increases. For further discussion on the relationship between the coupling and number of iterations, see~\cite{AWR}.

\subsubsection{Multi Zone Example}
\begin{figure}[htb]
    \begin{center}
   \subfigure[]{\includegraphics[scale=0.4]{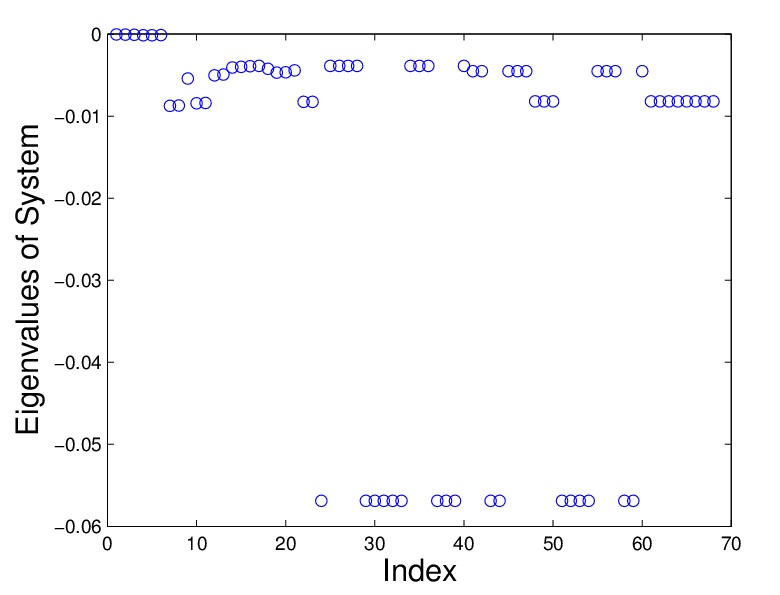}}
   \subfigure[]{\includegraphics[scale=0.4]{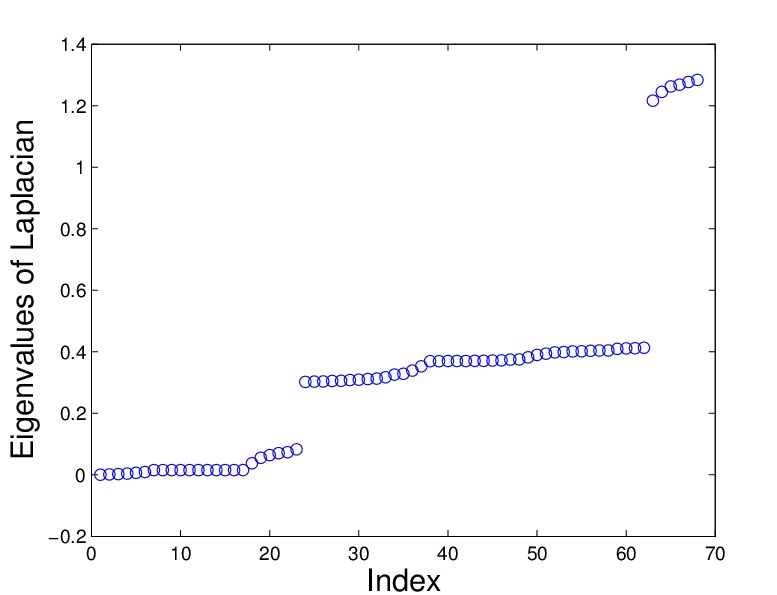}}
    \end{center}
    \caption{(a) Shows the there bands of eigenvalues of the time averaged $A(t;\xi)$ for nominal parameter values. (b) First spectral gap in graph Laplacian revealing $23$ subsystems in the network model.} \label{Fig:spectraldecomposition}
\end{figure}

In this section, we consider a larger $6$ zone building thermal network model with $68$ states. This model admits a decomposition into $23$ subsystems, as revealed by the spectral graph approach (see figure \ref{Fig:spectraldecomposition}b). This decomposition is consistent with three different time scales (associated with external and internal wall temperature, and internal zone temperatures) present in the system, as shown by the three bands in figure \ref{Fig:spectraldecomposition}a).

\begin{figure}[hbt]
\begin{center}
\subfigure[]{\includegraphics[scale=0.4]{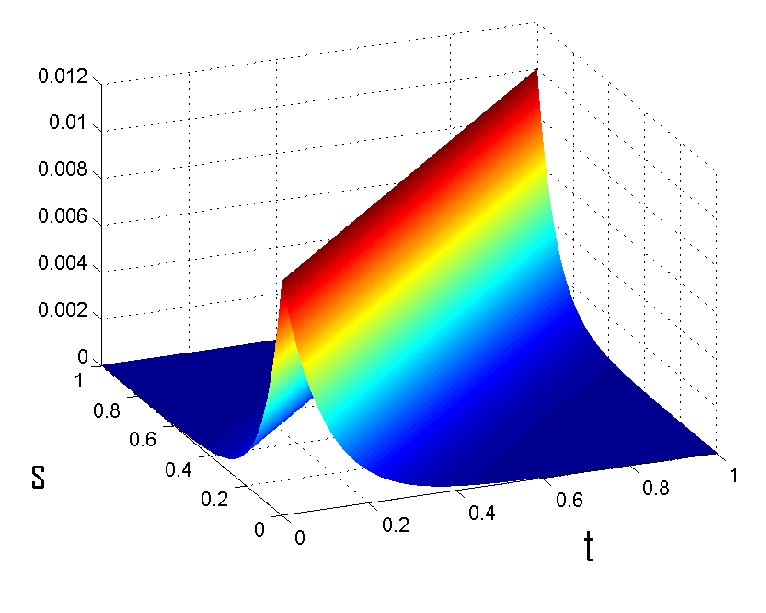}}
\subfigure[]{\includegraphics[scale=0.32]{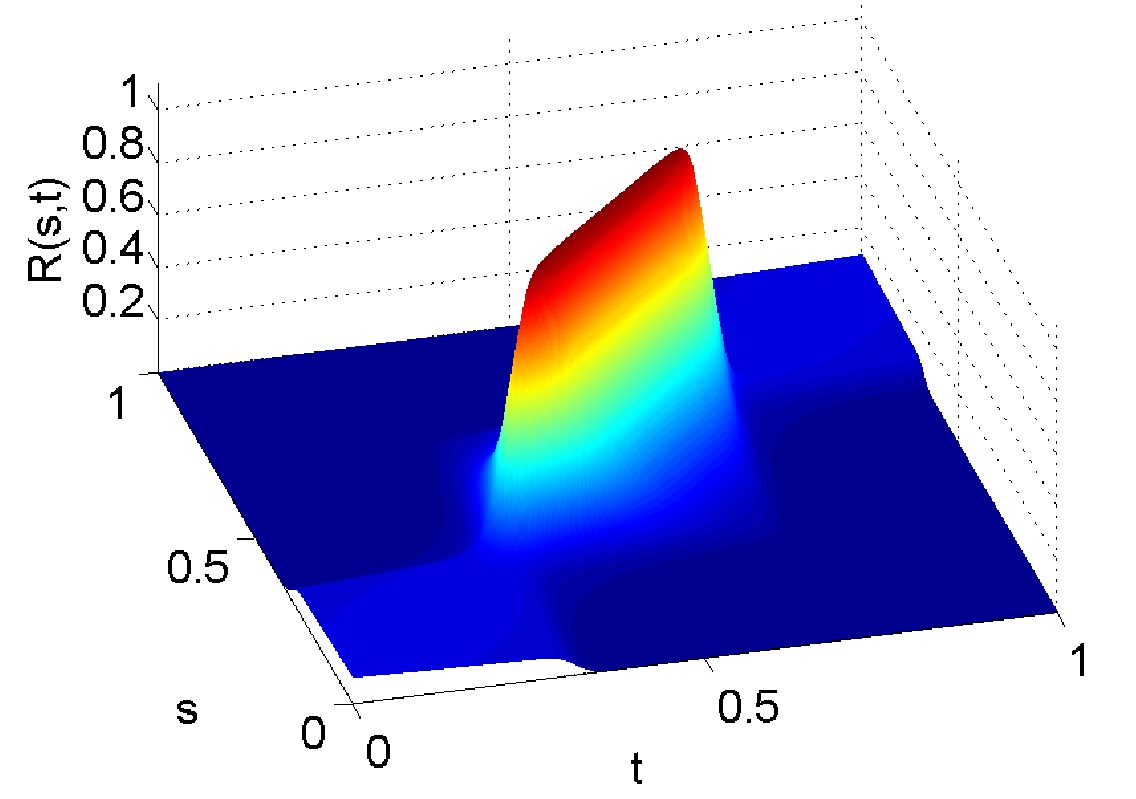}}
\end{center}
\caption{a) Covariance Kernel (\ref{covgauss1} for external load with $T_c=0.1$ and $\sigma=0.1$. b) Covariance kernel (\ref{covgauss2}) for internal load with $t_1=t_2=0.3$, $a=20$, $\sigma=0.1$.}\label{covplots}
\end{figure}

\begin{figure}[htb]
\begin{center}
\includegraphics[scale=0.5]{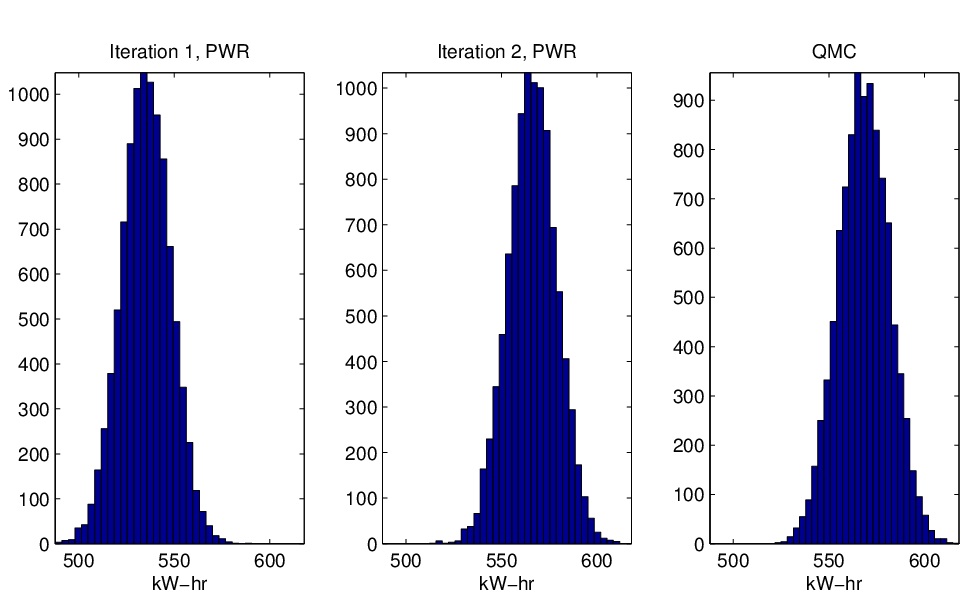}
\end{center}
\caption{Histogram of building energy computation for two iterations in PWR.
Also shown is the corresponding histogram obtained by QMC for comparison.}
\label{Fig:buildPWR}
\end{figure}

Next we demonstrate non-intrusive PWR approach to compute uncertainty in energy consumption due to both parametric uncertainty and time varying uncertain loads. As described earlier, we use KL expansion to transform time varying uncertainty into parametric form.

\paragraph{KL Expansion \cite{KL}:} Let $\{X_t=X(\xi,t),t\in[a,b]\}$ be a quadratic mean square second-order stochastic process with covariance functions $R(t,s)$. If $\{\phi_n(t)\}$ are the eigenfunctions of the integral operator with kernel $R(\cdot,\cdot)$ and $\{\lambda_n\}$ the corresponding eigenvalues, i.e.
\begin{equation}\label{kernel}
\int_{a}^bR(t,s)\phi_n(s)ds=\lambda_n\phi_n(t),\qquad t\in[a,b]
\end{equation}
then,
\begin{equation}\label{KL}
X(t,\theta)=\overline{X}(t)+\lim_{N\rightarrow\infty}\sum_{n=1}^N\sqrt{\lambda_n}a_n(\xi)\phi_n(t),\qquad \mbox{uniformaly for } t\in[a,b]
\end{equation}
where, $\overline{X}(t)$ is the mean of the process and the limit is taken in the quadratic mean sense. The random coefficients $\{a_n\}$  satisfy
\begin{equation}\label{coeff}
a_n(\theta)=\frac{1}{\sqrt{\lambda_n}}\int_{a}^b(X(\xi,t)-\overline{X}(t))\phi_n(t)dt
\end{equation}
and are uncorrelated $E[a_ma_n]=\delta_{mn}$. The basis functions  also satisfy orthogonality property
\begin{equation}\label{ucor}
\int_{a}^b\phi_m(t)\phi_n(t)dt=\delta_{mn},
\end{equation}
and the kernel admits an expansion of the form
\begin{equation}\label{KLkernle}
R(s,t)=\lim_{N\rightarrow\infty}\sum_{n=1}^N\lambda_n\phi_n(t)\phi_n(s).
\end{equation}
Generally, analytical solution to the eigenvalue problem (\ref{kernel}),  also known as Fredholm equation of second kind is not available. Several numerical techniques have been proposed, we used the expansion method described in \cite{expnKL}.

For applying KL expansion to the building problem, we assume that the stochastic disturbances $(T_{amb}(t),Q_{s}(t),Q_{int}(t))$ are Gaussian processes. This guarantees that the random variables $a_n$ in the KL expansion  are independent Gaussian random variables with a zero mean (\cite{expnKL}). Let the joint distribution of a nonstationary Gaussian process be,
\begin{equation}\label{jtgauss}
f(X_t,X_s)=\frac{1}{2\pi\sigma(s)\sigma(t)\sqrt{1-\rho(t,s)}}e^{-\frac{1}{2(1-\rho^2(t,s))}\left(\frac{x_t^2}{\sigma^2(t)}+\frac{x_s^2}{\sigma^2(s)}-\frac{2\rho(t,s)x_sx_t}{\sigma(t)\sigma(s)}\right)}
\end{equation}
where, $\rho(t,s)$ is the correlation coefficient and is related to covariance kernel as
\begin{equation}\label{covgauss}
R(t,s)=\rho(t,s)\sigma(t)\sigma(s).
\end{equation}
We assumed the processes  $T_{amb}(t),Q_{s}(t)$ to have a stationary  exponential correlation function
\begin{equation}\label{covgauss1}
R(t,s)=\sigma^2e^{-\frac{|t-s|}{T_c}},
\end{equation}
with a constant variance $\sigma^2$ and a constant correlation time scale $T_c$. For the internal occupancy load $Q_{int}(t)$ we constructed $R(t,s)$ as follows. For a typical office building, we know that the occupancy load is negligible with low variance during early and later parts of the day. On the other hand during peak hours in the middle of the day the occupant load can show significantly high variability. To capture this effect we divided the normalized time domain $[0,1]=[0,t_1]\cup(t_1,t_2)\cup(t_2,1)$ and obtained the desired variation by choosing (in expression \ref{covgauss})
\begin{equation}\label{covgauss2}
\sigma(t)=\sigma\left(\tanh(a(t-t_1))-\tanh(a(x-t_2))\right)/2,\quad \rho(t,s)=e^{-\frac{|t-s|}{T_c(s,t)}},
\end{equation}
with the correlation time scale
\begin{eqnarray}\label{Tc}
T_c(s,t)&=&(1-\tanh(a(t-t_1)))(1-\tanh(a(s-t_1)))/4\notag\\
& &+(1+\tanh(a(t-t_2)))(1+\tanh(a(s-t_2)))/4,
\end{eqnarray}
and parameter $a$ controls the slope of $\tanh$ function. Figure \ref{covplots} shows the covariance kernel for external $(T_{amb}(t),Q_{s}(t))$ and internal $Q_{int}(t)$ loads. For the choice of parameters indicated in the figure \ref{covplots}, we found using the expansion method \cite{expnKL} with Legendre polynomials as the basis functions, that KL expansion upto order $3$ and upto order $6$ can capture more that $90\%$ of total variance, for internal and external loads, respectively.

In UQ computation, we considered the effect of $14$ random variables comprising of external wall thermal resistance in the $6$ zones, and first dominant random variable obtained in the KL representation of internal load (for each zone) and first two dominant random variable obtained in KL expansion for solar load. Figure \ref{Fig:buildPWR} show the non-intrusive PWR results on the decomposed network model. As is evident, the iterations converge rapidly in two steps with a distribution close to that obtained from QMC (using a 25000-sample Sobol sequence) applied to the $68$ thermal network model (\ref{Buildmodel}) as a whole.

\begin{figure}[hbt]
\begin{center}
\includegraphics[scale=0.3]{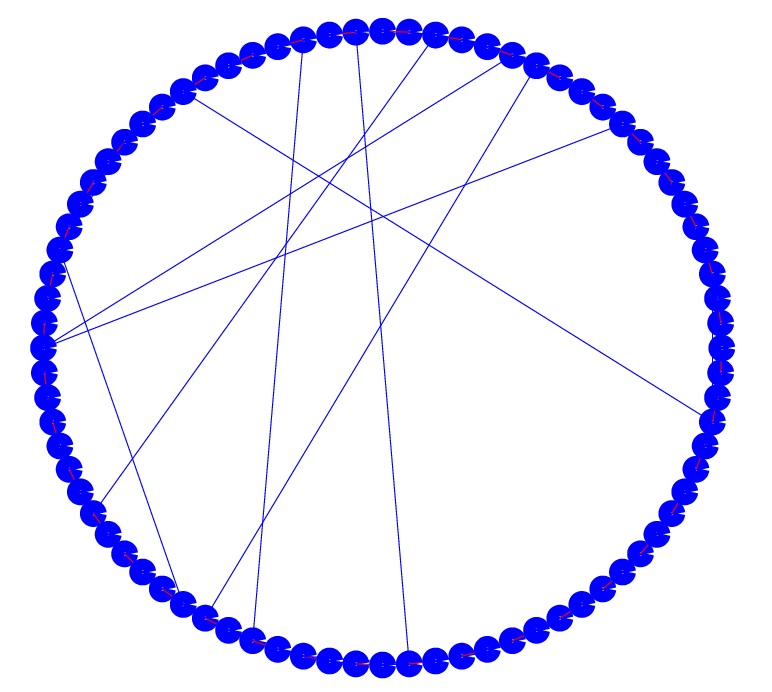}
\includegraphics[scale=0.3]{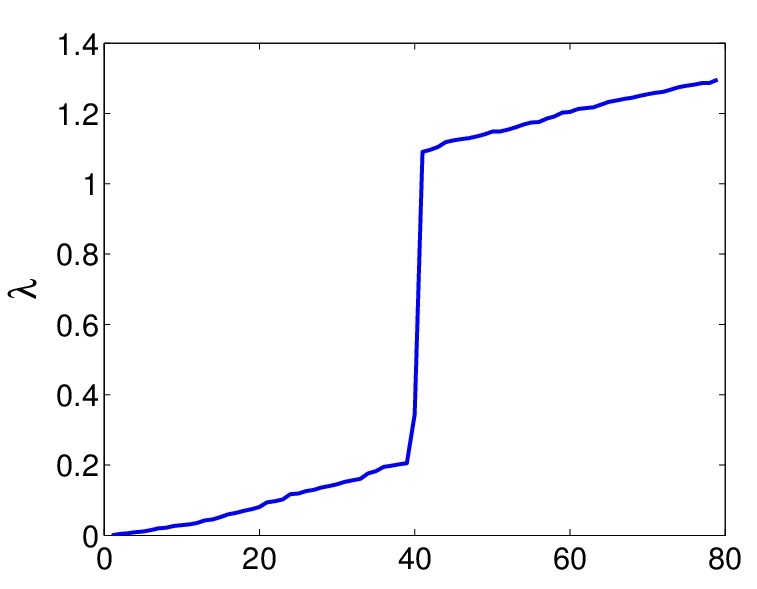}\\
\caption{Left panel shows a network  of $N=80$ phase only oscillators. Right panel shows spectral gap in eigenvalues of normalized graph Laplacian, that reveals that there are $40$ weakly interacting subsystems.}\label{fig1}
\end{center}
\end{figure}

\begin{figure}[hbt]
\begin{center}
\includegraphics[scale=0.4]{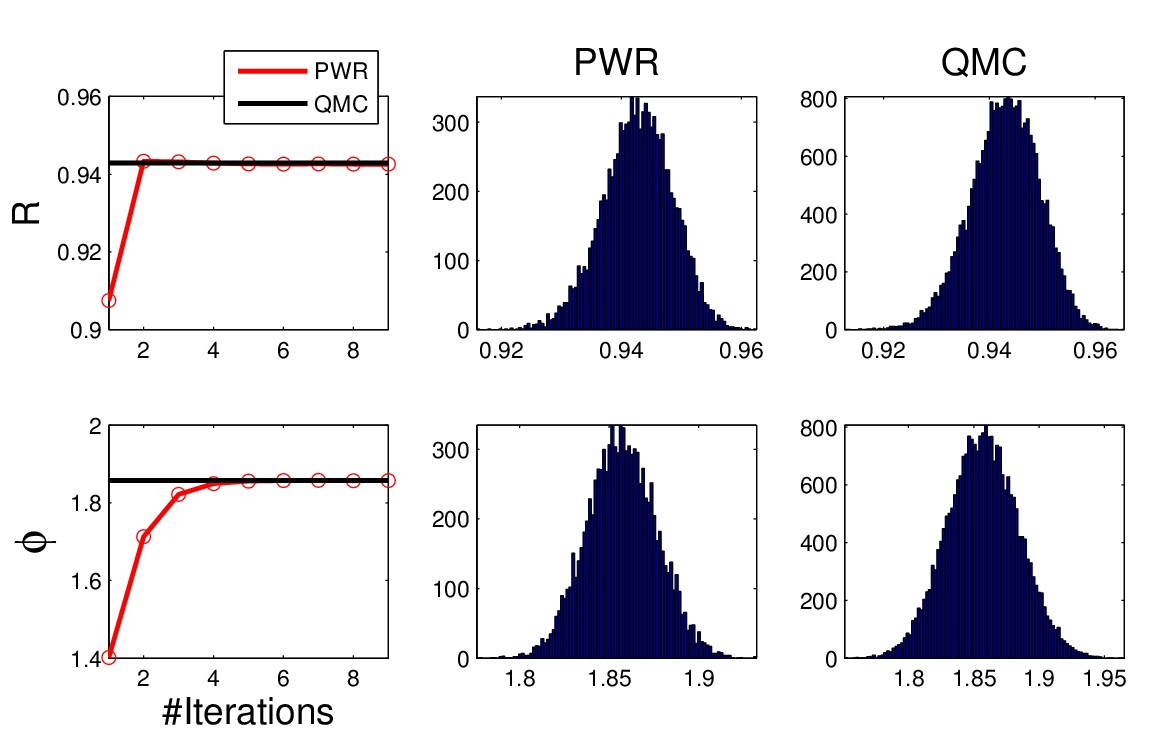}\\
\caption{Convergence of mean of the magnitude $R(t)$ and phase $\phi(t)$, and the respective histograms at $t=0.5$.}\label{fig2}
\end{center}
\end{figure}

\subsection{Coupled Oscillators}
Finally, we consider a coupled phase only oscillator system which is governed by nonlinear equations
\begin{equation}\label{ocslliator}
\dot{x}_i=\omega_i+\sum_{j=1}^N K_{ij} \sin(x_j-x_i),\qquad i=1,\cdots,n,
\end{equation}
where, $n=80$ is the number of oscillators, $\omega_i,i=1,\cdots,n$ is the angular frequency of oscillators and $K=[K_{ij}]$ is the coupling matrix. The frequencies $\omega_i$ of every alternative oscillator i.e. $i=1,3,\cdots,79$ is assumed to be uncertain with a Gaussian distribution with $20\%$ tolerance (i.e. with a total $p=40$ uncertain parameters); all the other parameters are assumed to take a fixed mean value. We are interested in the distribution of the synchronization parameters, $R(t)$ and phase $\phi(t)$, defined by $R(t)e^{\phi(t)}=\frac{1}{N}\sum_{j=1}^N e^{ix_j(t)}$. Figure \ref{fig1} shows the topology of the network of oscillators (left panel), along with the eigenvalue spectrum of the graph Laplacian (right panel). The spectral gap at $40$, implies $40$ weakly interacting subsystems in the network.

Figure \ref{fig2} shows UQ results obtained by application of non-intrusive PWR to the decomposed system with $l_s=5$, $l_c=2$. We make comparison with QMC, in which the complete system (\ref{ocslliator})  is solved at $25,000$  Sobol points \cite{Kuo1}. Remarkably the PWR converges in $4-5$ iterations giving very similar results to that of QMC. It would be infeasible to use full grid collocation for the networks as a  whole, since even with lowest level of collocation grid, i.e. $l=2$ for each parameter, the number of samples required become $\mathbf{R}_F=2^{40}=1.0995e+012!$.

\section{Conclusion and Future Work}\label{conc}
In this paper we have proposed an uncertainty quantification approach which exploits the underlying dynamics and structure of the system. In specific we considered a class of networked system, whose subsystems are dynamically weakly coupled to each other. We showed how these weak interactions can be exploited to overcome the dimensionality curse associated with traditional UQ methods. By integrating graph decomposition and waveform relaxation with generalized polynomial chaos and probabilistic collocation framework, we proposed an iterative UQ approach which we called \emph{probabilistic waveform relaxation}. We developed both intrusive and non-intrusive forms of PWR. We proved that this iterative scheme converges and illustrated it on several examples with promising results. Several questions need to be further investigated, these include: how the choice parameters associated with PWR algorithm affects its rate of convergence and the approximation error. In order to exploit multiple time scales that may be present in a system, multigrid extension \cite{mg} of PWR will be desirable.

\section{Acknowledgements}
This work was in part supported by DARPA DSO (Dr. Cindy Daniell PM) under AFOSR contract FA9550-07-C-0024 (Dr. Fariba Fahroo PM). Any opinions, findings and conclusions or recommendations expressed in this material are those of the author(s) and do not necessarily reflect the views of the AFOSR or DARPA.











\bibliographystyle{unsrt}

\bibliography{pwrbib}
\end{document}